\documentclass[twocolumn,amsmath]{aastex702}
\usepackage{multirow}
\usepackage{stackengine}
\usepackage{enumitem}
\usepackage[nosort]{cleveref}

\submitjournal{ApJ}
\begin{document}

\title{XRISM/Resolve Study of Large Scale Properties in Tycho's Supernova Remnant} 

\author[orcid=0009-0009-7793-9087]{Tyler E. Hanover}
\affiliation{Clemson University \\
Department of Physics and Astronomy \\
Clemson, SC 29634-0978, USA}
\email{thanove@clemson.edu}

\author[orcid=0009-0008-2949-862X]{Mark D. Leising}
\affiliation{Clemson University \\
Department of Physics and Astronomy \\
Clemson, SC 29634-0978, USA}
\email{lmark@clemson.edu}

\correspondingauthor{thanove@clemson.edu}

% ==========================================================================================
\begin{abstract}

Resolve, the microcalorimeter array aboard the %NASA/JAXA collaborative 
X-Ray Imaging and Spectroscopy Mission (XRISM), provides an unprecedented spectral view of Tycho's Supernova Remnant (SNR). % with its non-dispersive average energy resolution of 5--7 eV in the entire observing band.
We fit the X-ray spectra of the five public archival XRISM observations of the SNR with a multi-component non-equilibrium ionization plasma model assuming spherical expansion and a uniform abundance pattern.
For pointings away from the center of the object, we divide each into three spectra based on angular distance from the remnant's geometric center and fit all three simultaneously with the center-pointing observation.
We tie elemental abundances % (except Fe)
together across all model components. %in an attempt to find a unified set of abundances.
We find that pseudo-spherical expansion fits well the Tycho SNR in the regions where Resolve has observed, although iron appears slower on the approaching side than the receding side.
We also find that a close-to-uniform abundance pattern across the SNR is consistent with these data.

\end{abstract}


% ==========================================================================================
\section{Introduction} \label{sec:intro} %

Tycho's Supernova Remnant (SNR) is one of the best observed X-ray objects in the sky. 
It is the result of a thermonuclear supernova (Type Ia) as determined from its early light curve \citep{Baade45}, delayed echoes of its optical spectrum \citep{Krause08}, and its X-ray spectrum.

The commonly accepted Type Ia model is a binary system consisting of a carbon-oxygen white dwarf (WD), which explodes via thermonuclear runaway due to the inability of the electron-degenerate WD to expand and compensate for increased pressure from fusion-generated energy \citep{Arnett69}, caused by accretion from a companion main-sequence or giant star that has overflowed its Roche lobe. The details of the progenitor system and the nature of the trigger and propagation of the nuclear burning remain unclear.

Analysis of X-ray spectra has proven to be crucial to our understanding of SNRs as they reveal strong emission lines from the elements produced in the explosion when heated by  shocks from interaction of the ejecta with the ambient medium.
These elements include the intermediate-mass elements (IMEs), Si, S, Ar, and Ca, that are typically found together in space and post-shock conditions, and the iron-group elements (IGEs) which are collectively Cr, Mn, Fe, and Ni.

Astronomers have been collecting X-rays from the remnant of Tycho's Supernova  (Tycho, hereafter) since the rocket  experiment of \citet{Gorenstein70} which detected 250 counts from the object in the 1--10 keV band.
The first detection of the Fe K lines in Tycho was with the Mullard Space Science Laboratory proportional counter aboard Ariel-5 \citep{Davison76} and the first detailed X-ray spectrum of Tycho which showed the IME lines was captured by the solid-state spectrometer aboard the HEAO 2 (Einstein) Observatory \citep{Becker80}.
The latter was fit with two-temperature Raymond-Smith models; allowing the two components to have different abundances improves the fit to the spectrum and captures the Ca K line flux more accurately.
Between this result and a lower-resolution spectrum from %the sky survey predecessor observatory 
HEAO 1 \citep{Pravdo80}, Tycho is found to be ``far from equilibrium''.
\citet{Seward83}, using the Einstein data from above along with other observations, were the first to show clumpiness in Tycho.
They find roughly 400 bright clumps of material which they describe as non-equilibrium enriched ejecta clumps.
This spectrum was modeled, along with others, with a non-equilibrium model spectrum from a hydrodynamic calculation with elementally stratified ejecta \citep{Hamilton86}. 
Spectra obtained by the {\it Tenma} proportional counter clearly required non-equilibrium spectral models as well \citep{Tsunemi86}.

A number of later works investigated Tycho with multiple non-equilibrium ionization (NEI) models and methods for deriving the kinematics of the plasma ejecta.  
\citet{Vancura95} show that there is a clear observed contrast between the soft %(0.7--1.8 keV) 
and hard %(1.8--4.5 keV)
X-ray bands for multiple parameters % abundance, tau, kT
when fitting Tycho spectra from the Broad Band X-ray Telescope with NEI models (along with supporting imaging data from multiple instruments).
Using ASCA spectra, \citet{Hwang98_asca} found that separate NEI plasma model components, one for IMEs and one for IGEs, were needed. These components had distinctly different temperatures and ionization timescales.
\citet{Decourchelle01} present the first observation of Tycho with XMM-Newton in which they find Fe K and Fe L line emission peak at distinctly different radii.
All three of these studies corroborate the notion of at least some Fe in Tycho has physically different conditions compared to those of the IMEs.

\citet{SatoHughes17} use Gaussian line fits to identify small-scale blobs with large line-of-sight velocities. \citet{Williams17} include proper motions of small features and NEI model fits to derive three-dimensional motions of blobs of shocked ejecta; they conclude that overall the remnant, and therefore the supernova ejecta, is quite spherically symmetric. \citet{Godinaud2025} decompose spatial and spectral components of over two hundred small regions, identifying fast patches and more global north-south velocity asymmetry. In all of these works, the kinematics are derived from shifts and broadening of line features with modest energy resolution measurements. Only with NEI model fits are Doppler shifts separated from those due to varying ionization states.

In this work we analyze data from the 36-pixel microcalorimeter array {\it Resolve} aboard the X-Ray Imaging and Spectroscopy Mission (XRISM) \citep{XRISM}, which has excellent energy resolution for an imaging spectrometer. This presents a new opportunity for more direct determinations and separation of kinematics and degree of ionization. 

The {\it Resolve} data now publicly available cover over one-half of the remnant. We wish to fit these spectra along four center-to-edge directions with a minimal global model of the SNR. We know from the above studies that we require two NEI plasmas, one for iron-peak elements and one for intermediate mass elements. In the count spectra it is quite apparent that lines from the approaching and receding sides are resolved in the center but not near the edges, so we require a spatially varying Doppler shift. In practice, we use separate approaching and receding spherical shells for each plasma component, IME and IGE. We use a single set of elemental abundances, although we allow for iron also among the intermediate mass elements, and a single power-law. Even recognizing that small-scale variations in  line energies have been observed, we ask whether we can reasonably fit these early {\it Resolve} data with this relatively simple model, and what that tells us about the supernova and remnant.  

% ==========================================================================================
\section{XRISM/Resolve Data} \label{sec:data}
\setcounter{footnote}{0}

\begin{deluxetable*}{cccccc}[ht]
    \tablewidth{0pt}
    \tablecaption{Archival XRISM/Resolve observations of Tycho's Supernova Remnant used in this work.}
        \tablehead{
        \colhead{Observation Number} & \colhead{Location} & \colhead{Pointing (RA,DEC)} & \colhead{Roll Angle} & \colhead{Exposure Time} & \colhead{Observation Date} % & \colhead{Public Date} 
        }
    \startdata
        000160000 & NW     & (6.2326, 64.1676) & $226.37^{\circ}$ & 217.9 ks & 2024-02-02 \\ % & 2025-08-31 
        201118010 & NE     & (6.4080, 64.1704) &   $4.89^{\circ}$ & 159.5 ks & 2024-09-23 \\ % & 2025-10-05 
        300060010 & Center & (6.3339, 64.1385) & $219.36^{\circ}$ & 183.9 ks & 2024-03-02 \\ % & 2025-08-31 
        300062010 & North  & (6.3371, 64.1945) & $184.87^{\circ}$ &  53.5 ks & 2024-03-22 \\ % & 2025-08-31 
        300063010 & South  & (6.3338, 64.0914) & $183.69^{\circ}$ &  55.9 ks & 2024-03-23 \\ % & 2025-08-31 
    \enddata 
    \tablecomments{As of this work's completion, two observations have been taken but are not yet archived: 202098010 with pointing in the SW (public date 2026-12-31) and 203084010 with pointing in the SE (public date 2027-08-26).} % Observation 202037010 is scheduled with pointing in the SE but not yet taken
    \label{tab:obs_data}
\end{deluxetable*}

\begin{figure}[tb]
    \centering
    \includegraphics[width=\columnwidth]{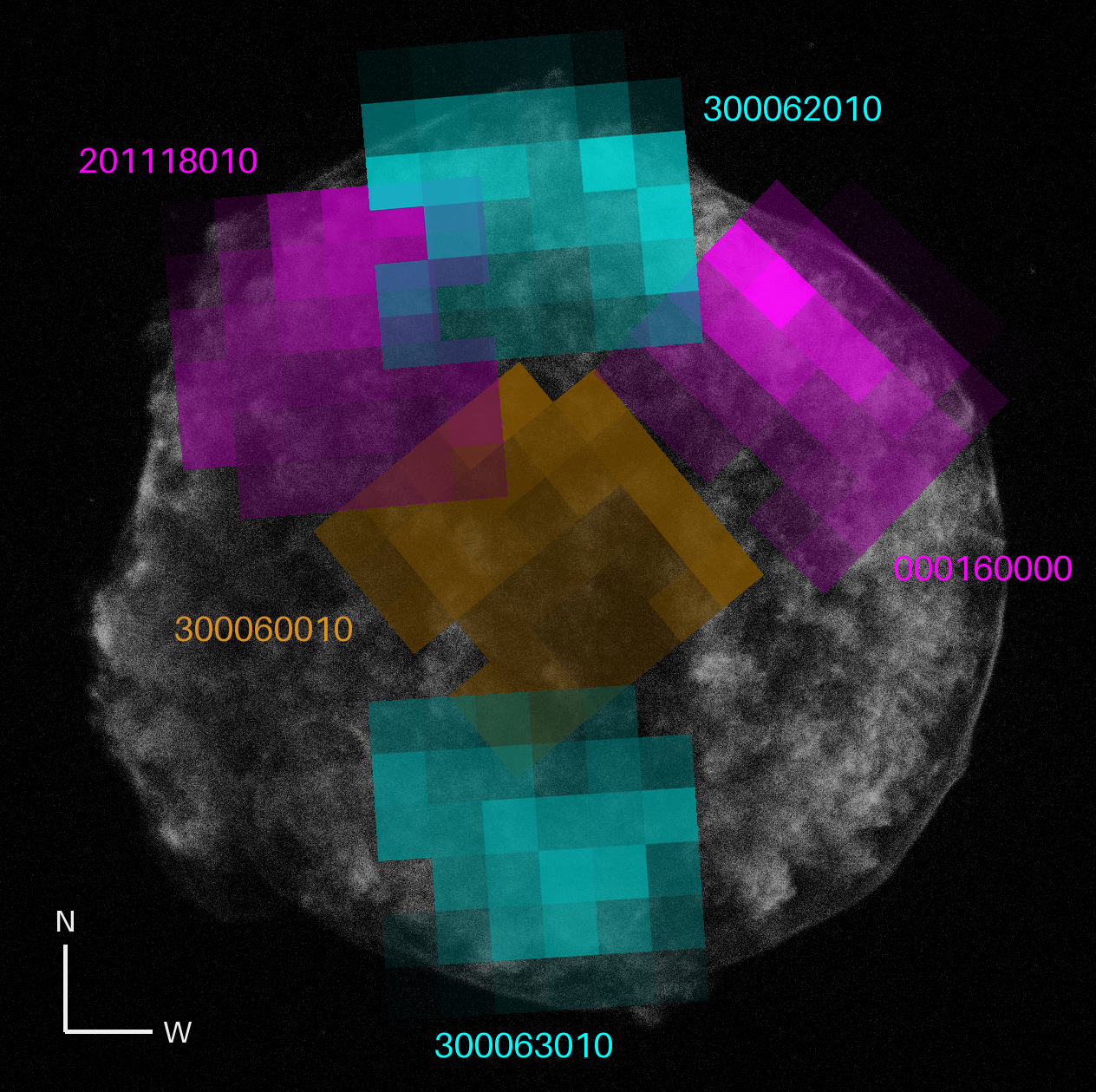}
    \caption{Here we show the {\it Resolve} detector images, normalized by counts s$^{-1}$, overlayed on a broadband image generated from multiple stacked Chandra/ACIS observations. The colors are purely for illustrative purposes to differentiate the data sets.}
    \label{fig:det_images}
\end{figure}

Resolve's new capability is a non-dispersive energy resolution of better than 7 eV in the entire observing band.
The calorimeters are kept near 50 mK via a liquid helium-based cooling system to achieve this. 
Radioactive $^{55}$Fe sources onboard provide energy and energy resolution calibration. %, which electron capture decay to $^{55}$Mn %that illuminate the array during portions of each observation.
For example, the {\it Resolve} energy scale report provided with the center-pointing observation shows the detector had an average FWHM for the Mn K$_\alpha$ line ($\sim\,$5.9 keV) of $4.53\pm0.02$ eV when  considering high-resolution primary events from the full array (excluding the dedicated calibration pixel).

The {\it Resolve} data used in this work, detailed in Table~\ref{tab:obs_data} and shown in \Cref{fig:det_images}, are acquired from NASA's HEASARC and processed with the following filters applied: %
\begin{itemize}[itemsep=0mm]
    \item High-resolution primary events (Grade 0) only %
    \item Pixel-pixel coincidence events screened out %
    \item Periods with high particle background excluded %(\texttt{"CORTIME.gt.8"})
    \item Off-pointing intervals excluded (necessary for spatially-resolved spectral analysis) % data outside \texttt{"ANG\_DIST.lt.0.1"}, 
    \item Pixel 12 (calibration only) and pixel 27 (shown to exhibit erratic gain jumps) excluded %
\end{itemize}

\noindent Spectra are then extracted from the filtered event files in XSELECT.
Redistribution matrix files (RMFs) are created using \texttt{rslmkrmf} with size set to X-large (\texttt{whichrmf=X}).
We use \texttt{splitrmf=yes} to processes the electron loss continuum separately with a coarse grid and \texttt{splitcomb=yes} to store it and the main RMF as two extensions in the same ($\approx500$ MB) FITS file.

\begin{figure}[hbt]
    \centering
    \includegraphics[width=\columnwidth]{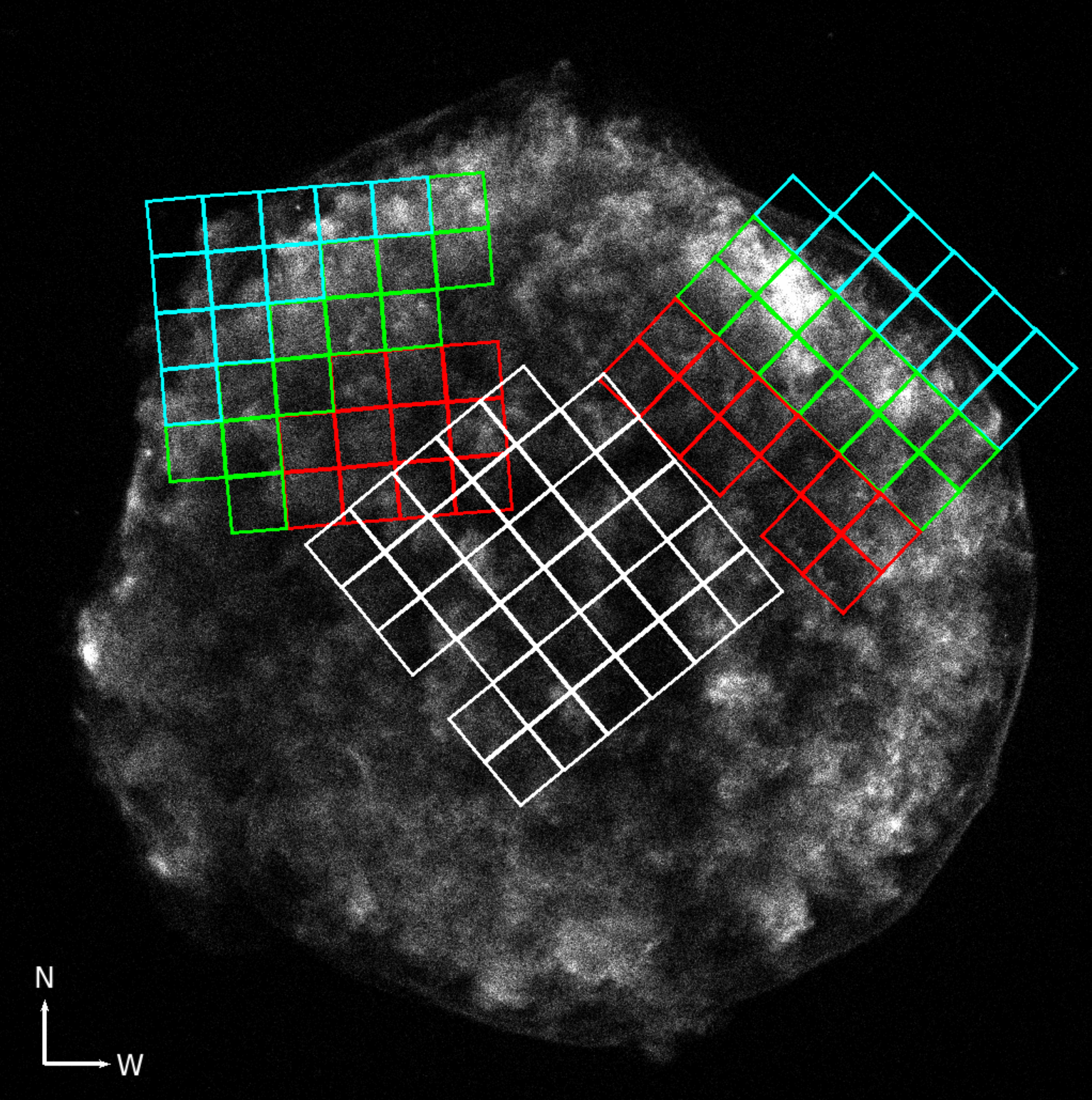}
    \includegraphics[width=\columnwidth]{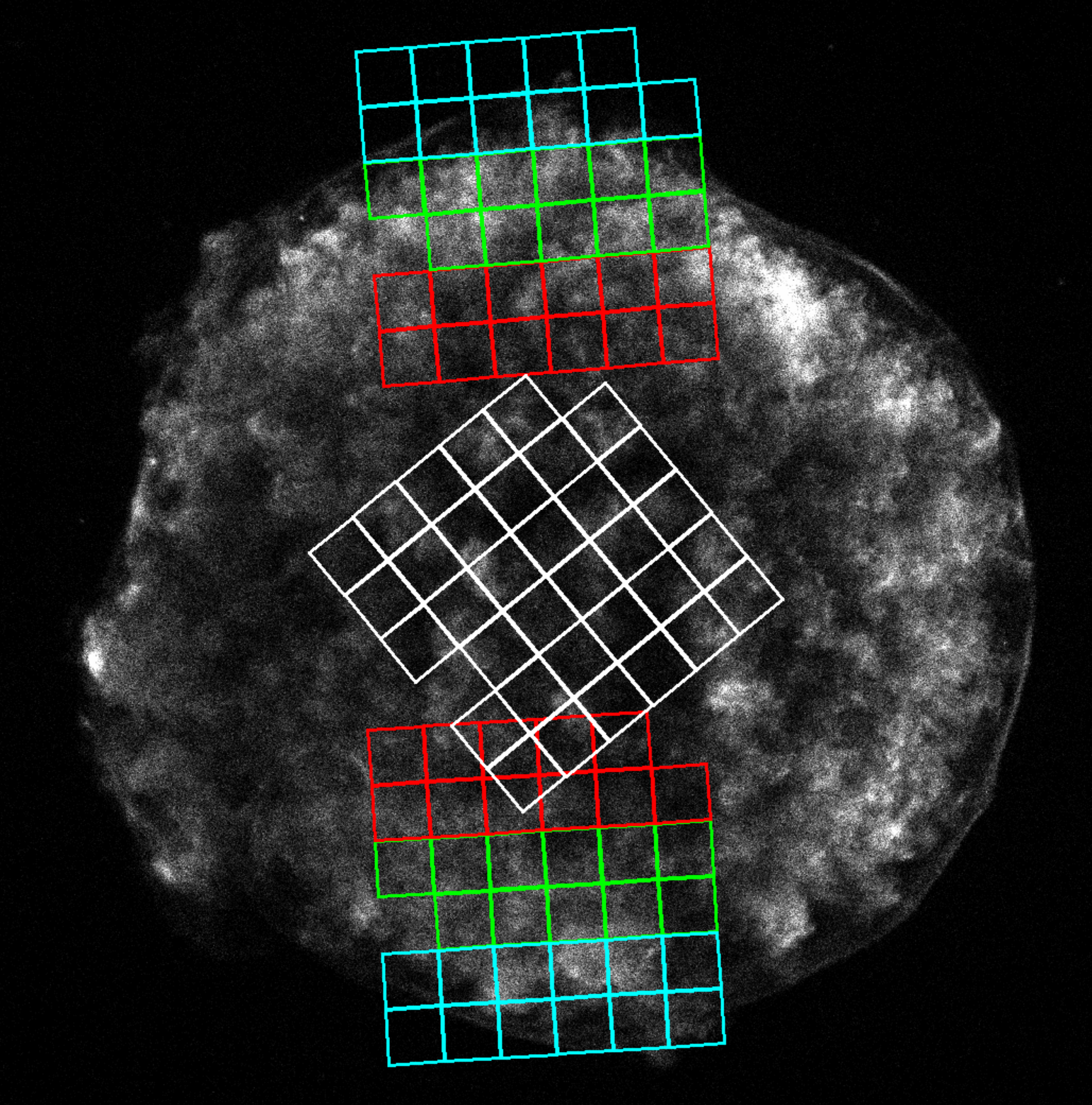}
    \caption{\textit{Top:} Angular breakdown of the {\it Resolve} array for observations 201118010 (NE) and 000160000 (NW), with observation 300060010 (Center) included in white, overlayed on the Chandra data from \Cref{fig:det_images}. \textit{Bottom:} The same figure but for observations 300062010 (North) and 300063010 (South).}
    \label{fig:radial_regions}
\end{figure}

The next step is generating an exposure map with \texttt{xaexpmap} to be used in ancillary response file (ARF) creation ---
this is done using the metatask \texttt{xaarfgen} which runs \texttt{xrtraytrace} and \texttt{xaxmaarfgen}.
As Tycho is an extended object, we run each ARF creation with \texttt{sourcetype=IMAGE} and feed it the background Chandra image\footnote{Chandra observations 23538, 24418, 24419, 24420, 24421, 25046, and 26426 from the same observing campaign (PI: B. Williams) are stacked to remove chip gaps.} used in \Cref{fig:det_images,fig:radial_regions}
with \texttt{numphoton=1000000} samples for the raytracing. We then set \texttt{source\_ra} and \texttt{source\_dec} to their respective values in Table~\ref{tab:obs_data}.

We generate {\it Resolve} backgrounds with \texttt{rslnxbgen}.
We fit the generated Non-X-Ray Background (NXB) with an empirical model provided by the XRISM team.
This background is not subtracted and will appear as a model component in our spectral fits.
As {\it Resolve}'s effective area is very small in the 13--15 keV band, we follow the example of \citet{Plucinsky2025} and use these data to normalize the NXB as they should be dominated by the particle background.

All {\it Resolve} events are binned in energy using the optimal binning algorithm of \citet{KaastraBleeker} in HEASoft.  For presentation, they are rebinned to $5\sigma$ with errors summed in quadrature. % \texttt{setplot rebin 5 1000}.

% ==========================================================================================
\section{Spectral Fitting} \label{sec:fitting} %

\begin{deluxetable*}{ccccc}[htb]
    \tablewidth{0pt}
    \tablecaption{Count-weighted geometric means, angular distances from the center, and resultant redshift scaling factors of the {\it Resolve} spectral regions shown in \Cref{fig:radial_regions}.}
    \tablehead{
        \colhead{Spectrum} & \colhead{Weighted Geometric} & \colhead{Angular Distance} & \colhead{Angular Distance} & \colhead{Redshift} \\[-2mm]
        \colhead{Name} & \colhead{Mean (RA,DEC)} & \colhead{to Mean} & \colhead{to Shock} & \colhead{Scaling}
        }
    \startdata
        NW Dim    & (6.2581, 64.1560) & $2.244'$ & $4.340'$ & 0.483 \\ 
        NW Bright & (6.2357, 64.1693) & $3.165'$ & $4.242'$ & 0.254 \\ 
        NW Shock  & (6.2100, 64.1786) & $4.037'$ & $4.216'$ & 0.043 \\ 
        \hline
        NE Dim    & (6.3835, 64.1585) & $1.769'$ & $4.318'$ & 0.590 \\ 
        NE Bright & (6.4031, 64.1742) & $2.806'$ & $4.240'$ & 0.338 \\ 
        NE Shock  & (6.4270, 64.1834) & $3.633'$ & $4.329'$ & 0.161 \\ 
        \hline
        North Dim    & (6.3305, 64.1782) & $2.385'$ & $4.613'$ & 0.483 \\ 
        North Bright & (6.3323, 64.1953) & $3.410'$ & $4.508'$ & 0.244 \\ 
        North Shock  & (6.3455, 64.2095) & $4.272'$ & $4.441'$ & 0.038 \\ 
        \hline
        South Dim    & (6.3430, 64.1073) & $1.885'$ & $4.159'$ & 0.547 \\ 
        South Bright & (6.3297, 64.0919) & $2.796'$ & $4.285'$ & 0.348 \\ 
        South Shock  & (6.3265, 64.0761) & $3.747'$ & $4.306'$ & 0.130 \\ 
        \hline
        NW Full    & (6.2366, 64.1675) & $3.084'$ & $4.264'$ & 0.277 \\ 
        NE Full    & (6.4054, 64.1728) & $2.782'$ & $4.333'$ & 0.358 \\ 
        North Full & (6.3349, 64.1932) & $3.284'$ & $4.424'$ & 0.258 \\ 
        South Full & (6.3332, 64.0929) & $2.734'$ & $4.270'$ & 0.360 \\ 
    \enddata
    \label{tab:geomeans}
    \tablecomments{Spectra naming is described in Section~\ref{sec:fitting}.}
\end{deluxetable*}

We seek the simplest model that can collectively explain these high-resolution spectra across the regions imaged. While Tycho is clearly not completely spherically symmetric, it is nearly circular on the sky and exhibits higher line-of-sight velocities near its center than edges, which is apparent in {\it Resolve} count spectra. So, we assume expanding spherical shells for each plasma component and map the expected line-of-sight velocity to the angular distance of each detector from the center. For each model component, the maximum expansion velocity (that of the center observation) is a free parameter.
When describing these fits, we mean angular distance from the center of Tycho when using the shorthand term ``radial''.

Spectra are fit in XSPEC \citep{XSPEC} mainly with the plane-parallel collisional plasma model \texttt{pshock} \citep{Borkowski2001}.
This is an NEI model which has the following free parameters: temperature $kT$, ionization timescale $\tau=n_e t$, redshift $z$, and a normalization parameter.\footnote{See \texttt{nei} and \texttt{pshock} models in the \href{https://heasarc.gsfc.nasa.gov/docs/software/xspec/manual/}{XSPEC user manual} for further details.}
Specifically, we use the \texttt{bvvpshock} variant which allows one to fit Gaussian velocity broadening $\sigma_v$ and individual elemental abundances for Z $\le30$.
We add a power-law to represent synchrotron emission and convolve the entire model with the Tuebingen-Boulder ISM absorption model \texttt{TBabs} \citep{Wilms2000}.
We fix hydrogen column density $n_\text{H}=10^{22}$ cm$^{-2}$, which allows for some internal absorption that varies throughout the remnant \citep{Katsuda2010}. % average I believe was 0.72e22

We use the hydrogen-poor plasma self-consistent electron density calculation  \citep{Leahy2024} implemented in XSPEC v12.15.0, as is appropriate for a Type Ia reverse shock. This requires the command \texttt{xset APECDOECOR yes}, and necessitates using finite difference derivatives for Levenberg-Marquardt minimization in v13.0.
% v13.0 release cannot use its new closed-form analytic derivative method
% We also considered using the new {\it nest} method, a multi-ellipsoidal nested sampler, but found it was too computationally expensive.

We fix the abundances of H, He, and N at 0 and all other non-varying abundances at 1 (i.e., solar). 
We therefore assume a solar abundance of carbon and normalize the abundances to that.
All NEI components in the fits below have the same abundance pattern, with exception of Fe, across all spectra in order to attempt to establish a global set of abundances
(e.g., Si$_1$ = Si$_2$ = Si$_3$... with only Si$_1$ varying).

Tying all abundances and NEI model parameters, including line-of-sight velocities (``redshift'' parameter in the models), across all model components and measurements yields decidedly unacceptable fits in C-Stat and in the residuals around line features. As we relax these constraints, we find that varying $kT$, $\tau$, and $\sigma_v$ for each measured spectrum, and splitting each plasma component into approaching and receding shells with separate redshift parameters, we can achieve reasonable fits, even with abundances tied across all components. 

Only the full center observation, which is in every fit, has a fitted redshift parameter; the rest of the spectral shifts are tied to that as constant fractions of that shift from count-weighted geometric means of the {\it Resolve} pixel positions. %assuming the pointing of the center observation is the geometric center of Tycho.
These redshift scaling factors are found in Table~\ref{tab:geomeans}.
The shock angular distances used to determine the redshift scaling factors are found by drawing lines from the pointing of the center observation through the geometric means of the other datasets to the shock positions as determined by a smoothed contour around Tycho.
This contour is at 3 counts which is widely considered the threshold for detection in Chandra data. % citation?
 
A visual breakdown of the {\it Resolve} sub-regions from Table~\ref{tab:geomeans} is shown in \Cref{fig:radial_regions}; as useful shorthand names only, the colored radial regions in this figure are referred to as {\bf ``Dim''} (red/inner), {\bf ``Bright''} (green/middle), and {\bf ``Shock''} (blue/outer) hereafter. 
%These names were chosen for the NW observation and were kept when creating the other sub-regions for consistency.
The North, South, and NW observations are oriented such that they are divided into three $2\times6$ pixel groups (pixels 12 and 27 are then removed) that progress roughly radially outward.  The NE observation was divided into three groups based on individual detector angular distances. 
The center observation is fit as a single spectrum.
  
All spectra are fit from 1.5--15.0 keV with the combined model \texttt{TBabs(4*bvvpshock + powerlaw) + NXB} where two \texttt{pshock} components are meant to represent  redshifted and blueshifted IME-dominant plasmas, respectively, and two are for IGE-dominant plasmas, with independent redshifts.

\renewcommand{\arraystretch}{1.2}
\begin{deluxetable*}{cccccc}[!htbp]
    \tablewidth{0pt}
    \tablecaption{Center observation \texttt{pshock} component redshifts in units of $10^{-3}$. % with 90\% MCMC errors. 
    All fits include the full center spectrum and the column headings are which additional spectra are fit simultaneously with it.} % change colors due to colorblind accomidations?? not sure if dashed and solid would work with the complexity of the model
    \tablehead{
        \colhead{Component} & \colhead{All Data} & \colhead{NW Radial} & \colhead{NE Radial} & \colhead{North Radial} & \colhead{South Radial}
    }
    \startdata
        IME Orange & $13.34_{-0.58}^{+0.48}$ & $15.31_{-0.66}^{+0.39}$ & $13.68_{-1.47}^{+0.76}$ & $13.56_{-2.87}^{+0.38}$ & $13.95_{-2.20}^{+0.76}$ \\
        IME Cyan & $-12.26_{-0.37}^{+0.39}$ & $-12.89_{-0.18}^{+0.61}$ & $-13.08_{-0.24}^{+0.36}$ & $-12.80_{-0.23}^{+0.62}$ & $-12.58_{-0.47}^{+0.49}$ \\
        IGE Red & $9.87_{-0.44}^{+0.88}$ & $13.02_{-0.41}^{+0.58}$ & $12.34_{-2.25}^{+1.59}$ & $13.39_{-0.66}^{+1.12}$ & $11.39_{-0.16}^{+2.41}$ \\
        IGE Blue & $-7.34_{-0.39}^{+0.33}$ & $-7.76_{-0.29}^{+0.45}$ & $-7.75_{-0.32}^{+0.49}$ & $-7.44_{-0.49}^{+0.20}$ & $-7.60_{-0.31}^{+0.38}$ \\
        \hline
        \multirow{2}{*}{C-Stat/d.o.f.} & 15672/14522 & 12112/11416 & 11280/11172 & 9943/10732 & 9683/10690 \\[-1mm]
        & (1.079) & (1.061) & (1.010) & (0.926) & (0.906) \\ % not sure where to put these C-Stats but here is fine
    \enddata
    \tablecomments{The leftmost column refers to the color of the components in \Cref{fig:NE_individual_spectra}. See \S4.1 for more about ``All Data''.}
    \label{tab:redshifts}
\end{deluxetable*}

Each \texttt{bvvpshock} component has thirteen free parameters ($kT$, eight abundances, $\tau$, $z$, $\sigma_v$, and norm).
When fitting four {\it Resolve} spectra simultaneously, each having four NEI components, power-law index and norm, and NXB amplitude, we have potentially 220 free parameters.
Tying the redshifts together would bring this down to 208. % and the large number of varying abundances would not be very informative.
By tying the abundances of the outer spectra to those of the center spectrum, we reduce the number of free parameters to 112.
Constraining the abundances of the IME and IGE plasma components to a single set (except for Fe) reduces the number of free parameters to 89.
We find that a single power-law index is adequate for all spectra, therefore the final number of free parameters for a simultaneous fit of four spectra is 86.
So in all spectra that are not the center-pointing one, the only parameters that are varying are power-law norm, NXB amplitude, and four each of $kT$, $\tau$, $\sigma_v$, and norm$_\text{pshock}$.
NXB amplitude is parameter 1 in the provided empirical model, which is a varying constant factor.

We calculate errors with the Markov chain Monte Carlo (MCMC) method in XSPEC, 
using the Goodman-Weare algorithm \citep{GoodmanWeare} with 400 walkers each talking 10,000 steps after a minimum of 5,000  burn-in steps.
The Goodman-Weare algorithm evenly divides the total number of steps among the walkers, so this is entered into XSPEC as \texttt{chain walkers 400}, \texttt{chain length 4000000}, and \texttt{chain burn 2000000}.
%The new Hamiltonian Monte Carlo method {\it hmc} in v13.0 is unavailable to us as it refuses to launch, which is expected per the XSPEC manual with a model like ours.
All errors are 90\% confidence unless stated otherwise.

These spectral fits and their MCMC runs take a significant amount of time due to each of them having over 10,000 degrees of freedom (d.o.f.) and over 80 free parameters.
Using our local computing cluster, fitting the {\it Resolve} spectra simultaneously took a few hours on 32 cores and the MCMC runs took up to a week on 100 cores.
%Thankfully, in our experience, XSPEC's built-in parallelization appears to scale better than logarithmic with an increasing number of tasks for Levenberg-Marquardt and Goodman-Weare.

% ==========================================================================================
\section{Fitting Results} \label{sec:results} 

\begin{figure*}[htb] 
    \centering
    \stackunder[3pt]{(a) NW Obs 000160000}{\includegraphics[width=0.49\textwidth]{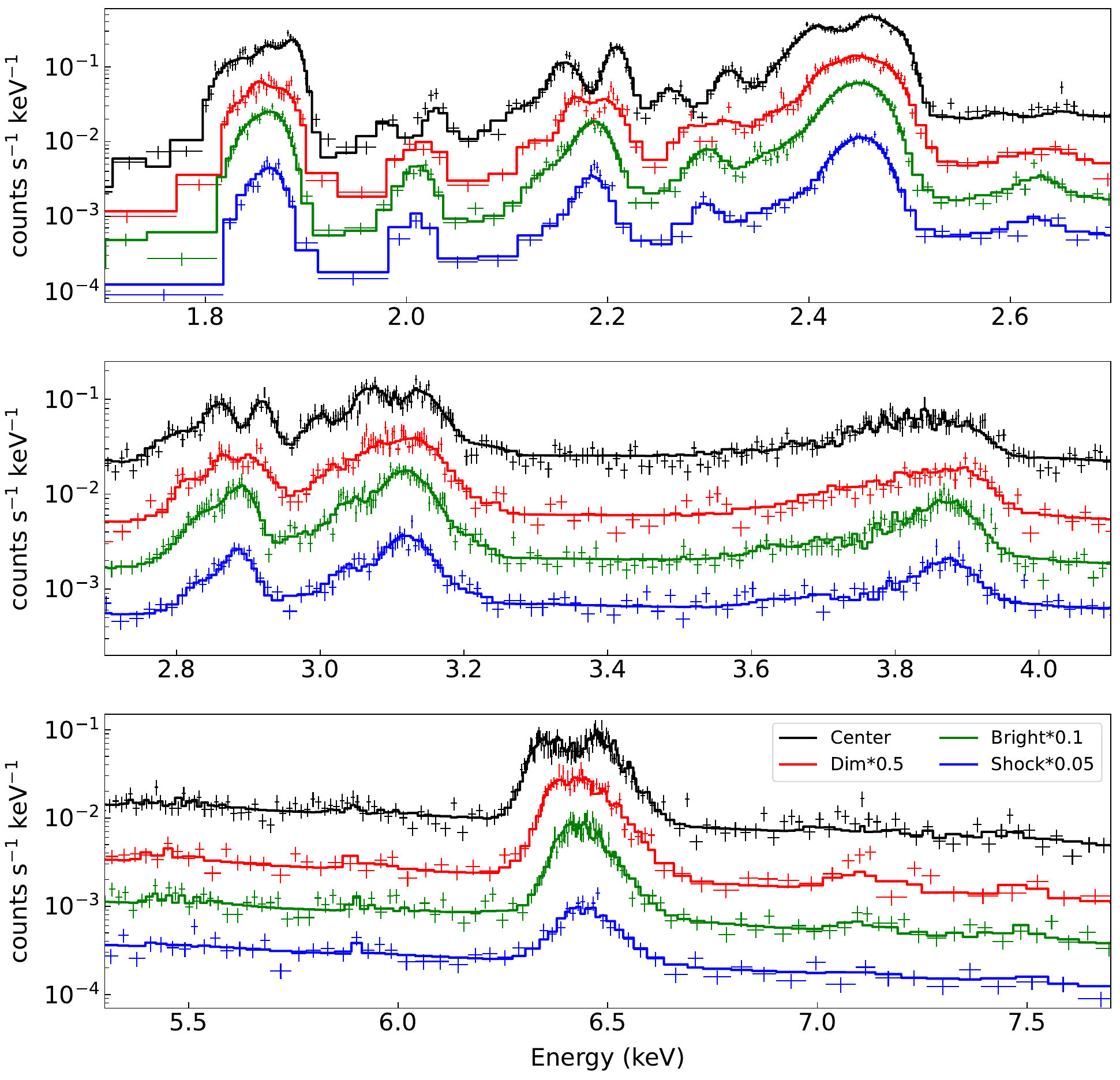}}%
    \hfill%
    \stackunder[3pt]{(b) NE Obs 201118010}{\includegraphics[width=0.49\textwidth]{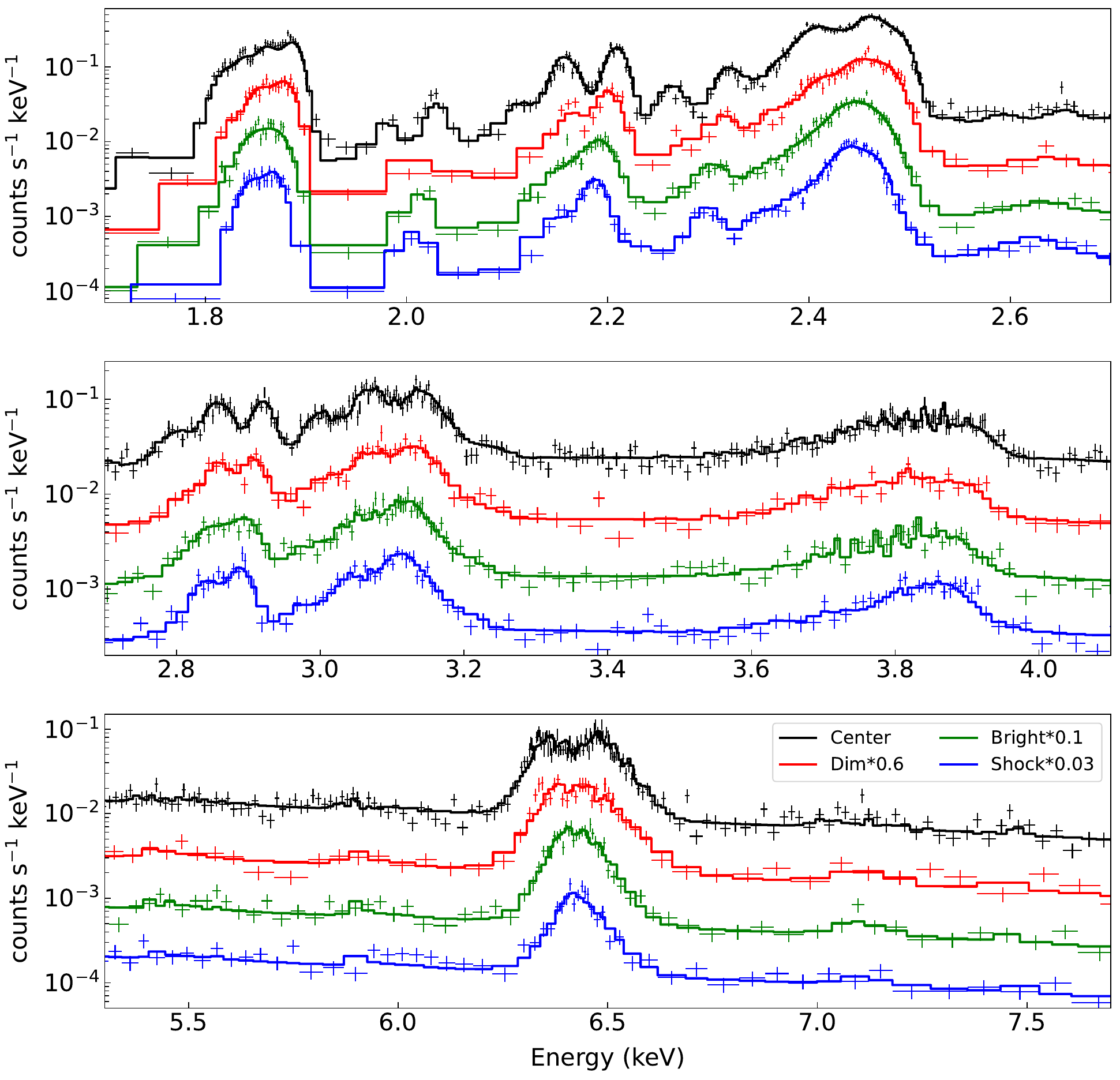}}%
    \vspace{10pt}%
    \hfill%
    \stackunder[3pt]{(c) North Obs 300062010}{\includegraphics[width=0.49\textwidth]{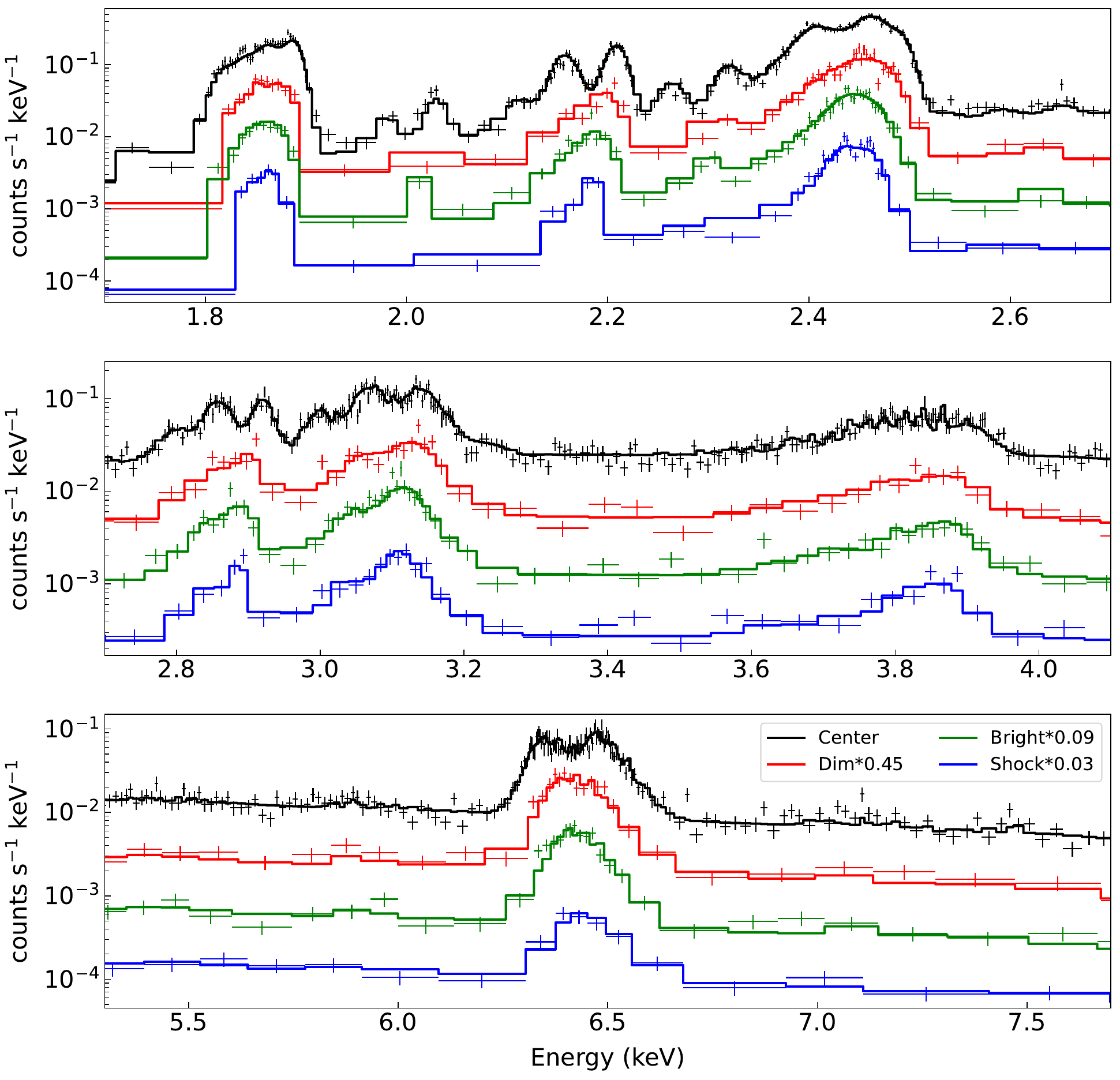}}%
    \hfill%
    \stackunder[3pt]{(d) South Obs 300063010}{\includegraphics[width=0.49\textwidth]{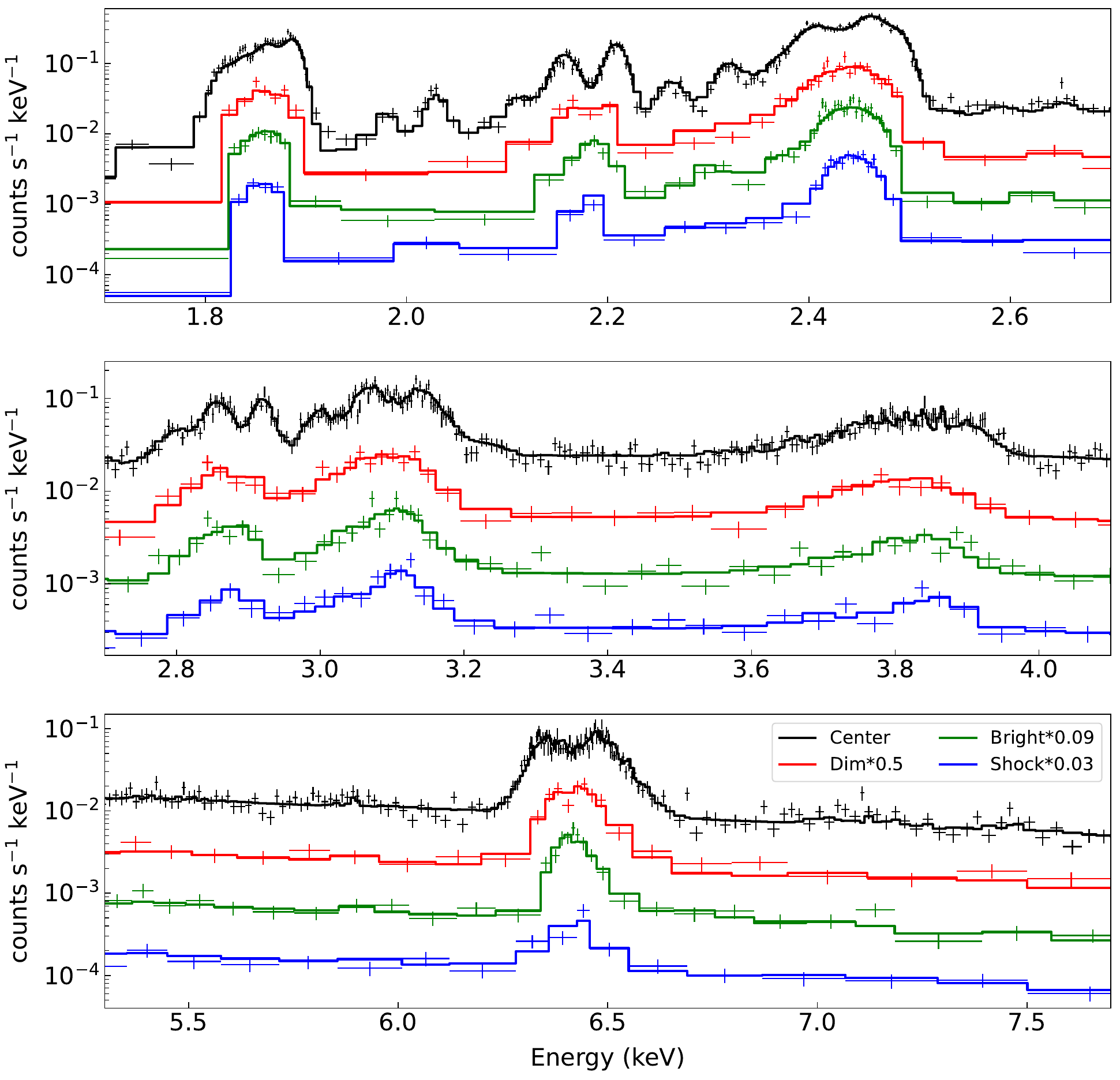}}%
    \caption{The four simultaneous center-edge spectral fits with three of the spectra in each scaled down for presentation. The spectra colors roughly match those of the regions in \Cref{fig:radial_regions}. In each, all abundances and redshifts are linked to those of the center spectrum (top of each plot) as described in the text. Three sections of each fit are shown from 1.7--2.7 keV, 2.7--4.1 keV, and 5.3--7.7 keV. We do not show the 4.1--5.3 keV band as it is continuum-dominated and above 7.7 keV we find no significant non-NXB emission lines.} 
    \label{fig:radial_fits}
\end{figure*}

\begin{figure*}[htb] 
    \centering
    \includegraphics[width=\textwidth]{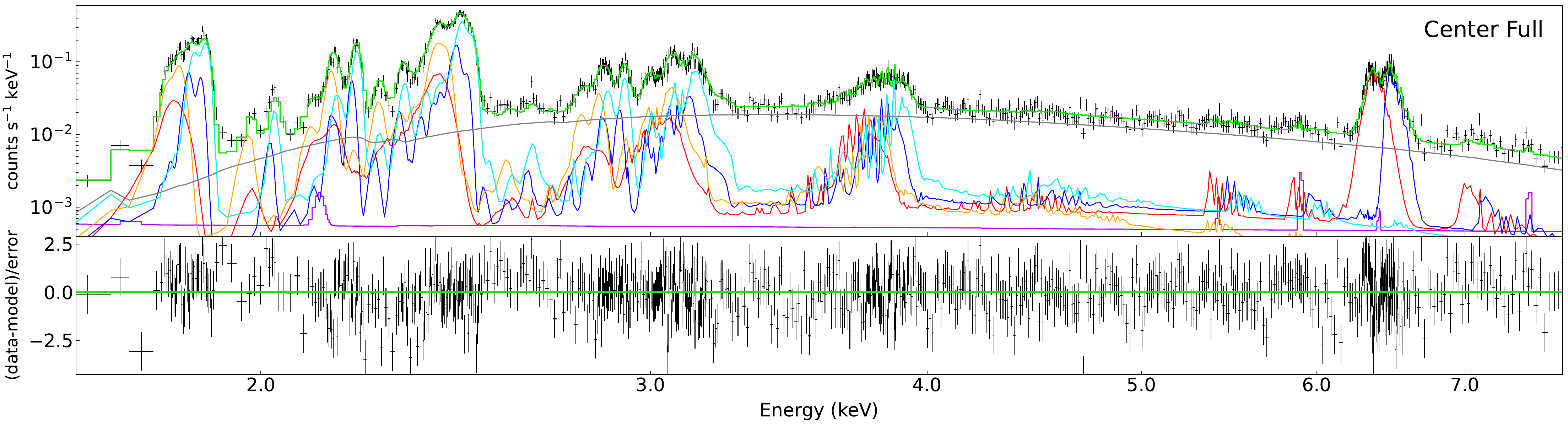}
    \includegraphics[width=\textwidth]{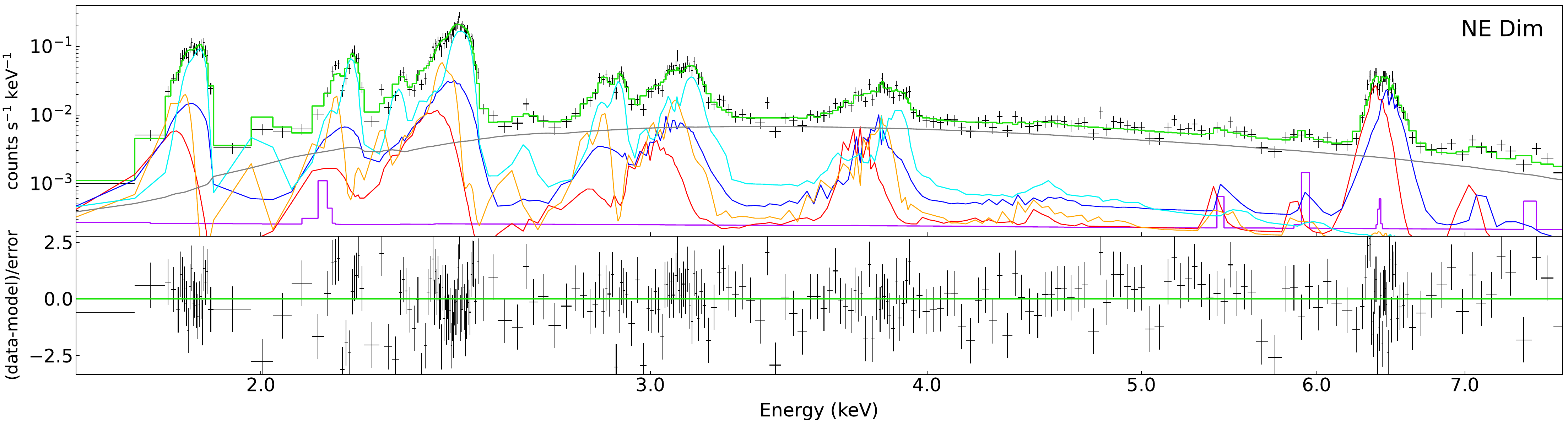}
    \includegraphics[width=\textwidth]{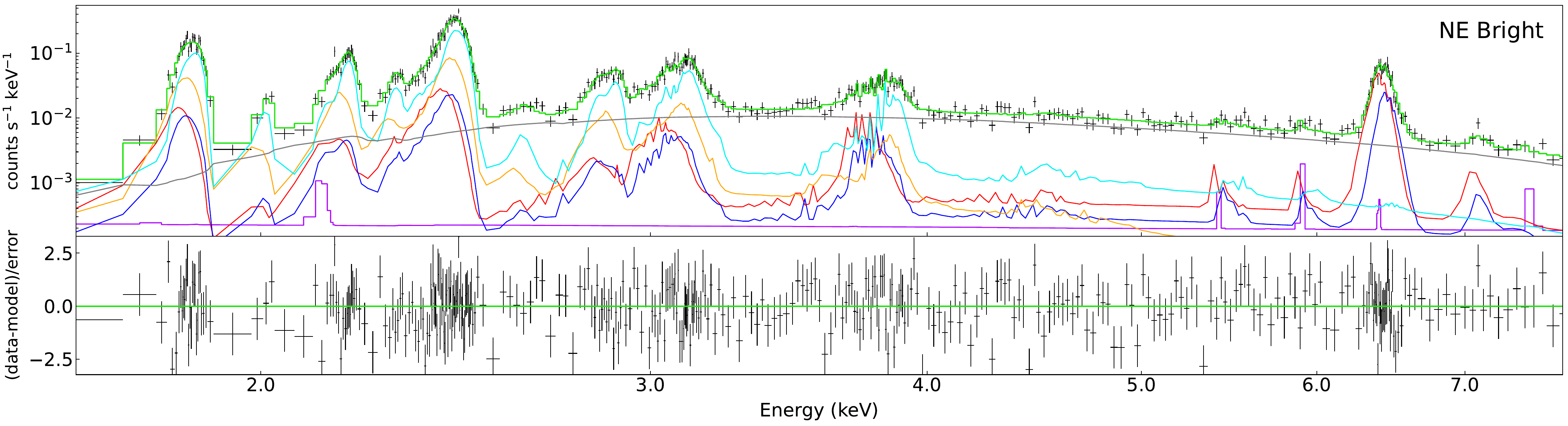}
    \includegraphics[width=\textwidth]{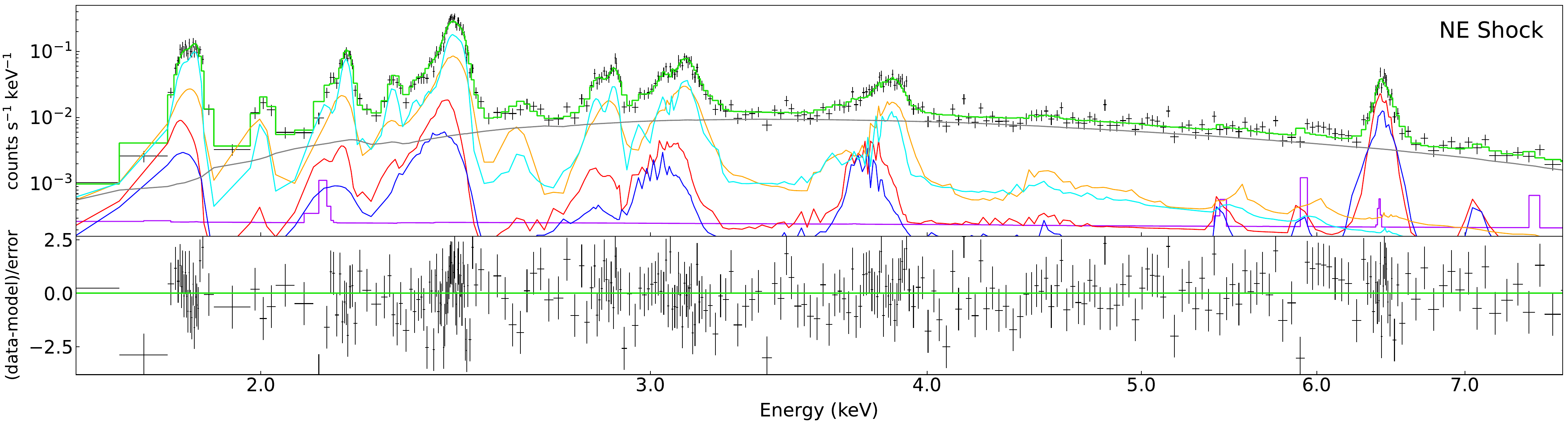}
    \caption{Each of the spectra from the NE simultaneous angular fit in panel (b) of \Cref{fig:radial_fits} %, plotted from 1.7--7.7 keV, 
    shown with their residuals. The model components depicted are as follows: red and blue are the IGE-dominant redshifted and blueshifted \texttt{pshock}, orange and cyan are the IME-dominant redshifted and blueshifted \texttt{pshock}, grey is the power-law, and purple is the NXB.}
    \label{fig:NE_individual_spectra}
\end{figure*}

\begin{deluxetable*}{cccccc}[pbth]
    \tablewidth{0pt}
    \tablecaption{Mass ratios of the abundances (see Appendix~\ref{appx:xspec}) from our simultaneous spectral fits. They are computed by taking the ratio of (XSPEC abundance)*(Solar abundance)*(atomic mass) % for the two elements 
    and carrying the errors forward in quadrature.}%
    \tablehead{
        \colhead{Ratio} & \colhead{All Data} & \colhead{NW Radial} & \colhead{NE Radial} & \colhead{North Radial} & \colhead{South Radial}
    }
    \startdata
        S / Si & $0.62_{-0.05}^{+0.05}$ & $0.61_{-0.12}^{+0.13}$ & $0.57_{-0.05}^{+0.05}$ & $0.56_{-0.09}^{+0.10}$ & $0.55_{-0.08}^{+0.08}$ \\
        % S / Si & $0.619_{-0.049}^{+0.051}$ & $0.614_{-0.124}^{+0.127}$ & $0.571_{-0.052}^{+0.050}$ & $0.560_{-0.093}^{+0.098}$ & $0.548_{-0.078}^{+0.076}$ \\
        Ar / Si & $0.148_{-0.013}^{+0.014}$ & $0.146_{-0.030}^{+0.031}$ & $0.119_{-0.013}^{+0.012}$ & $0.121_{-0.020}^{+0.024}$ & $0.109_{-0.017}^{+0.018}$ \\
        Ca / Si & $0.159_{-0.014}^{+0.015}$ & $0.149_{-0.032}^{+0.031}$ & $0.115_{-0.014}^{+0.012}$ & $0.104_{-0.016}^{+0.024}$ & $0.096_{-0.017}^{+0.016}$ \\
        \hline
        Cr / Fe$_{\text{\,IGE}}$ & $0.013_{-0.004}^{+0.004}$ & $0.015_{-0.007}^{+0.006}$ & $0.012_{-0.004}^{+0.006}$ & $0.008_{-0.004}^{+0.011}$ & $0.007_{-0.002}^{+0.002}$ \\
        Mn / Fe$_{\text{\,IGE}}$ & $0.011_{-0.003}^{+0.005}$ & $0.009_{-0.006}^{+0.006}$ & $0.010_{-0.005}^{+0.006}$ & $0.011_{-0.008}^{+0.006}$ & $0.005_{-0.004}^{+0.004}$ \\
        Ni / Fe$_{\text{\,IGE}}$ & $0.005_{-0.003}^{+0.002}$ & $0.011_{-0.008}^{+0.007}$ & $\le0.009$ & $0.006_{-0.003}^{+0.009}$ & $0.008_{-0.004}^{+0.007}$ \\
    \enddata
    \tablecomments{Solar abundances from \citet{Wilms2000}}
    \label{tab:mass_ratios}
\end{deluxetable*}

\begin{figure}[htb]
    \centering
    \includegraphics[width=\columnwidth]{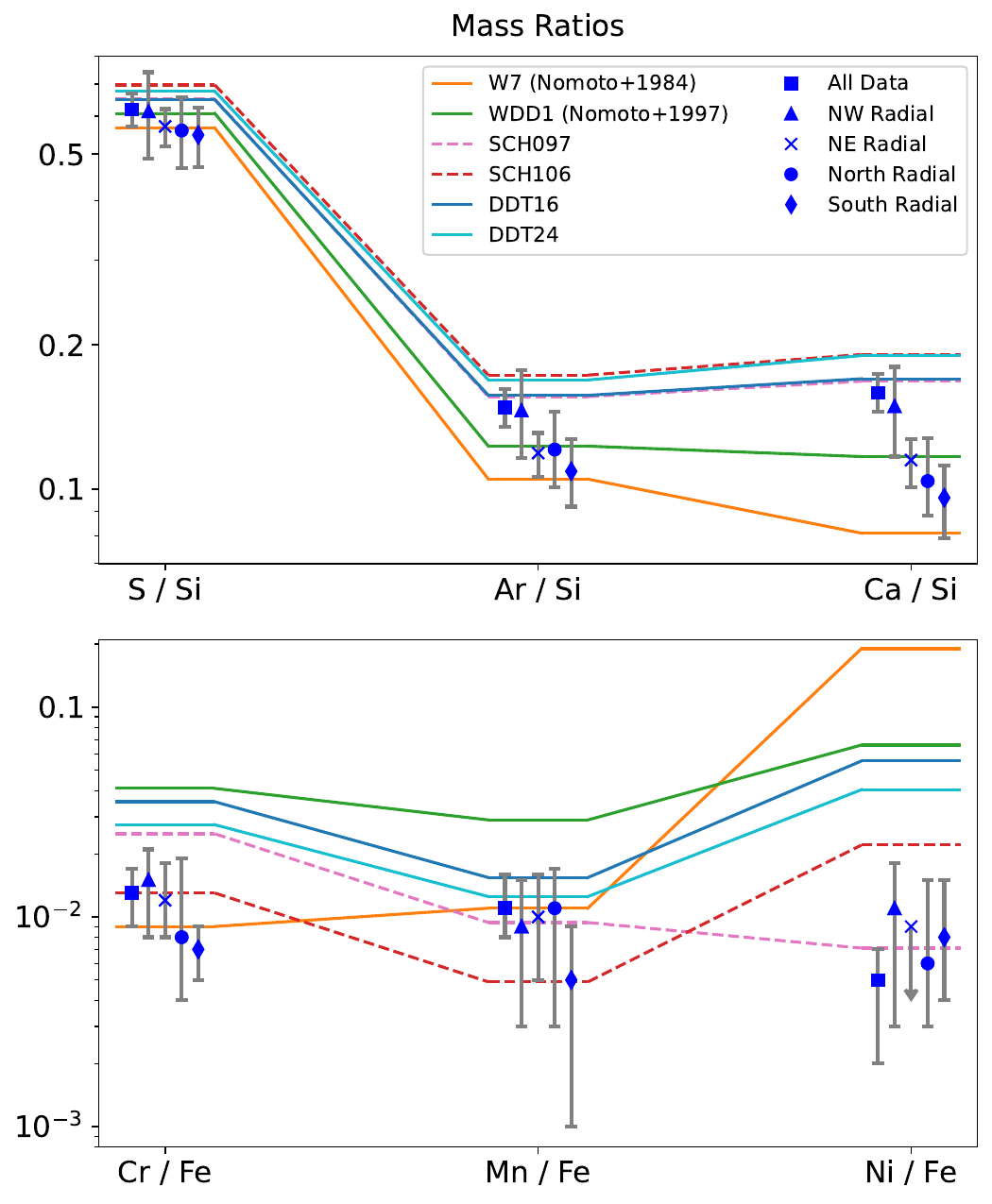}
    \caption{Mass ratios from Table~\ref{tab:mass_ratios} (blue data points with gray error bars) plotted against ratios from some of the works referenced in Section~\ref{sec:discussion} (the various solid/dashed lines). The latter four models are taken from \citet{Martinez-Rodriguez18}.} 
    \label{fig:mass_ratios}
\end{figure}

Fits of the center spectrum with four \texttt{bvvpshock} components do not give vastly improved C-Stat over those with only three, but the latter show poor 
fits to details of the Si K$_\alpha$ and Fe K$_\alpha$ features, with significant runs of residuals.
After achieving a good fit of the center spectrum, we add the outer spectra and refit separately for each direction with the constraints described in Section~\ref{sec:fitting}.
We list the final center spectrum redshifts from each fit in Table~\ref{tab:redshifts} and the raw XSPEC abundances in Appendix~\ref{appx:xspec}.

In \Cref{fig:radial_fits}, we show all four simultaneous center-to-edge fits. 
We show the individual spectra with their model components and residuals for the NE Radial fit in \Cref{fig:NE_individual_spectra} as an example.
We choose to represent the redshifted components as orange (IME) and red (IGE) while the blueshifted components are represented as cyan (IME) and blue (IGE).
We use this color scheme when referencing the individual model components.
The~fit parameters for the NE Radial fit are tabulated in Appendix~\ref{appx:xspec}.

The fits are not exceptional everywhere in each spectrum, but they are reasonable. 
With all of the constraints in place, the largest C-Stat/d.o.f. of the center-to-edge fits is 12112/11416 (1.061) for the NW direction. In all fits, the center spectrum contributes most to C-Stat, probably because its lines have the most complicated structure.
%Degrees of freedom can only truly be calculated for the entire fit though due to the many parameters tied between the spectra.
As a test, we perform a fit for the NW where all abundances are free, with only redshifts and power-law indices tied, and it reduced C-Stat/d.o.f. from 1.061 to 1.055 (11918/11297), a marginal improvement. 
We also perform a test fit where the IMEs and IGEs are completely separated; this fit has the same number of d.o.f. and reduced C-Stat goes up by 4\%.
Completely independent treatment of the elemental groups is quite common as the Fe K line emission peaks at a smaller radius than the Si K line emission \citep[e.g.,][]{Decourchelle01}.
So one might expect an improvement when fully separating the elemental groups. However, we do not find that to be the case. %It is reasonable to assume that some of the shocked IME ejecta has the same conditions as the currently shocked IGE ejecta. 
% IGE components may fill in fluctuations in the data as they contribute to the IME lines

%the small bump near 1.7 keV %which has a slightly different appearance depending on the spectrum and visual binning, is an instrumental effect and the strange shape of the Si K$_\alpha$ complex %'s low-energy side is due to poor effective area below 2 keV (as {\it Resolve}'s gate valve is currently in the closed position).
Some of the line features have quite interesting shapes, the S K$_\alpha$ (mix of Li-like and He-like) has a low-energy shoulder which is likely due to higher energy Si transitions such as Ly$_\beta$.
The Ar K$_\beta$ + Ca K$_\alpha$ feature, spanning roughly 3.6--4.0 keV, is quite rich, forming a pseudo-continuum. 
In lower energy features belonging to Si, S, and Ar, there are some distinguishable ionization states.
For example, Si XII (Li-like), XIII (He-like), and XIV (H-like) dominate the line emission in the 1.8--2.1 keV band.
%One can see, particularly in the Fe K$_\alpha$ complex, how line-of-sight velocity of the SNR decreases with distance from the center, as the double peaks merge.

An interesting result is our model shows a number of K-shell transitions from $n>6$ for sulfur XV (He-like) in the center spectrum's IME Cyan component. 
This feature appears at 3.2--3.3 keV as a shoulder on  the Ar K$_\alpha$ line. 
Using \texttt{return\_linelist} from PyAtomDB \citep{PyAtomDB} 
we expect lines out to $n=29$ ($\approx3.223$ keV) in an NEI plasma with $kT$ and $\tau$ of the IME Cyan component in, e.g., Figure~\ref{fig:NE_individual_spectra}'s top panel. 
The combination of the 1.3\% blueshift and {\it Resolve's} resolution make their presence apparent.
%Generally one would not see any of these high-$n$ K-shell lines in an astrophysical spectrum, although they are always present in a plasma with these conditions, being that they are quite weak compared to K$_\alpha$ and K$_\beta$.

The spectra are all fit with four NEI components, but some of the simpler spectra formally require fewer than four, e.g., the NW Shock where a good fit is obtainable with only two components; in the simultaneous fit, the IGE Blue component is completely suppressed as is the IME Orange component. 
However, we keep all components in the combined fits for consistency.
We fit each set of spectra multiple times with different initial conditions to try and stop bias from initial conditions and the fits we present are all tested this way. 
%We do note an underfitting of the Mn K$_\alpha$ feature in some fits, which is most likely due to the global elemental abundance pattern used in our model.
%The lines may be suppressed due to there being little-to-no flux in some model components.

Among all the simultaneous fits shown in this work (including the one in \S4.1), only 11 out of the 448 free parameters have multimodal posteriors.
Several of them occur because local minima are present within the $3\sigma$ confidence interval. This is not unexpected as the complexity of the parameter space creates the possibility of more local minima. If these parameters were to move into those minima, it would not change the conclusions of this work. Some other parameters are poorly determined, 
mostly due to low norms in their components, but they are not multimodal. %As redshifts get smaller going radially outward, the components can also become degenerate which leads to more local minima as well.

The four center-to-edge fits are somewhat similar, but of course have their differences. NE and South do not capture the shock as independently as NW and North, simply due to the pointing of XRISM during the observations. However, all four still exhibit the same behavior of the double peaks merging the further out one goes from the center of Tycho. Although each spectrum has quite a number of aspects that could be discussed, below we briefly discuss some interesting features in the fits. 

The NW fit's blueshifted Si Ly$_\alpha$ ($\sim\,$2.05 keV) is more underfit in its center spectrum compared to the other radial fits. 
It is possible that the differences in plasma conditions and/or overall brightness between the NW portion of Tycho and the other observed regions cause this effect. Another feature not fit well is the NW Dim Fe K$_\beta$ ($\sim\,$7.1 keV) which appears to be a double peak separated more than predicted. Something like this, which also appears at the $\sim\,$2.2 keV Si He$_\beta$ feature in NW Dim, can potentially be attributed to {\it Resolve}'s $1.3'$ half-power diameter being larger than the individual $0.5'$ detectors. Even though this would cause the NW Dim pixels to capture more photons from closer to the shock, it would also cause knots further interior to that spectrum's region to be more influential to the spectrum than expected, which could explain the line separation.

The NE radial fit has higher abundance values which are most likely a consequence of the norms being lower overall as their ratios remain consistent with those in the other three radial fits. Splitting up the detector pixels by radius, instead of evenly like in the other three outer datasets, should not be a significant factor as each sub-region still contains either eleven or twelve pixels. This is the only fit where the nickel abundance was an upper limit; it is quite possible the 7.48 keV NXB Ni K$_\alpha$ line could be masking some of the weak source line flux. The Si Ly$_\alpha$ feature almost vanishes in NE Dim but is clearly present in the other three spectra. This may be due to broadening, redshifts, or the spectral binning.

The North and South radial fits have less counts in them due to lower exposure times, but still agree somewhat well with the others in terms of redshifts and abundance ratios (which will be discussed in the next section). Both of the ``Shock'' spectra for these fits capture at least some of the forward shock, but the North fit's $\sim\,$2.9 keV (likely S He$_\beta$) is much sharper than the South's. The same goes for the Ar K$_\beta$ + Ca K$_\alpha$ feature which may be due to binning, or the South shock may have weaker lines compared to the North. The South Shock spectrum has a higher redshift scaling factor than the North Shock spectrum, so lines that appear weak may be the result of the double peaks being slightly more separated with less photons being collected compared to the NW and NE.

\subsection{Fitting of All Data} \label{subsec:alldata} 

We fit all five full {\it Resolve} datasets at once as test of the model to see if it will still fit well over a majority of the remnant (while also introducing an additional $\sim\,$4000 bins).
We include this fit as ``All Data'' in \Cref{fig:mass_ratios} and multiple tables.
The number of free parameters increases from 86 to 104 for this fit.
C-Stat/d.o.f. = 1.079 is very satisfactory in this case as there are 14,522 d.o.f. when the all of the currently available {\it Resolve} observations are fit together.

Its mass ratios, which are discussed below, are slightly larger overall compared to the radial fits and seems to be more in agreement with NW than the others. 
It is possible due to the large number of counts in the NW observation, from higher brightness and longer exposure time, that the All Data fit is skewed towards the properties of the NW. 
The other three radial fits agree very well with each other, but not always with the All Data fit.
The IGE Red component's shift is also quite different from the center-to-edge fits, but this is most likely due to the larger number of counts in the Fe lines from using the full datasets of the other four observations.
% It is possible our model, although it fits the spectra well, may need some tweaking

% ==========================================================================================
\section{Discussion} \label{sec:discussion}

We find that we can simultaneously fit reasonably well all {\it Resolve} Tycho data, divided into thirteen spectra, with four plasmas each (approaching and receding for both IGE and IME) with abundances tied across them, and Doppler shifts tied within each element group. We find similar fit parameters for four different azimuthal directions fit separately, although all include the same center observation.
The fit statistics are not highly improbable and we do not find heavily 
structured residuals. At the current level of signal-to-noise, Tycho appears to be described by roughly spherically symmetric expansion (on scales of $\geq$ 3 arcmin or 2 pc) 
and a close-to-uniform set of elemental abundances. Because we find significant broadening of all spectral components, this does not mean the spherical shells are necessarily thin. 

Mass ratios resulting from our fits are in Table~\ref{tab:mass_ratios}.
IME and IGE ratios are separated; global ratios across the groups are not meaningful, as all of the IGE ejecta is likely not yet shocked. % reference?
The measured ratios are generally consistent among different fits, although a few vary outside 90\% confidence limits.

These elemental ratios can be compared to total ejecta ratios of Type Ia models. This is valid for iron-peak ratios, if those ratios do not vary much from inner, unshocked regions, to the outer already shocked iron-rich regions. 
We compare to the models W7  of \citet{Nomoto1984} and  WDD1  of \citet{Nomoto1997}, which are modified deflagration and delayed detonation models, respectively. We also compare to models of \citet{Martinez-Rodriguez18}, including two delayed detonation and two sub-Chandrasekhar-mass models. These are shown in Figure~\ref{fig:mass_ratios}. The measured IME ratios do not discriminate significantly among these models. The Cr/Fe and Ni/Fe ratios measured favor the SCH097 and SCH106 sub-M$_{\text{Ch}}$ models, and formally exclude these particular M$_{\text{Ch}}$ models.

In our fits, we find the blueshifted shell of the IGE emission is slower than the redshifted portion.
We can see this directly in the Fe K$_\alpha$ line.
For example, when fit with two Gaussians, we find the peaks of the two sides of the Fe K$_\alpha$ complex in the center observation are at $6340_{-5}^{+5}$ eV and $6469_{-7}^{+6}$ eV (90\% errors with $\Delta$C-Stat).
%% forcing z, tau, etc doesn't work
When we force the IGE redshifts and blueshifts to be equal in a trial fit of NEI models, we find the fit quality decreases significantly; to compensate for the IGE Blue component's lower shift, the IGE Red component is somewhat suppressed, the blue component itself is considerably broadened, and the Fe K complex cannot be fit without runs of significant residuals.

\citet{SatoHughes17} find the mean velocities of their blueshifted and redshifted Chandra spectral regions to be $-3220\pm970$ km s$^{-1}$ and $4980\pm740$ km s$^{-1}$, respectively.
This corresponds to $z=(-10.73\pm3.23)\times10^{-3}$ and $z=(16.60\pm2.47)\times10^{-3}$, respectively, which is consistent with all of our IGE blueshifts but with only the North Radial IGE redshift within 90\%.
They also suggest the front face may be on average denser by a factor of $\lesssim \sqrt{3}$ than the back face.

\citet{Fujimaru26} calculated several 3D Type Ia models and their interactions with external media, producing spectra at various times for comparison with {\it Resolve} data. Their Si and Fe lines, for example from narrow directions through the centers of the full remnants, are generally broader than the features we see here and lack clear double peaks. This might be because they find a wide range of expansion speeds of blobs along a given line-of-sight through the model remnants or because they obtain larger thermal broadening than actually occurs in Tycho. In our fits, 
the one-sigma broadening, which captures all of a shell's spread in radial expansion velocity, non-radial bulk motions, and thermal broadening, is typically one-half or less of the redshift in IME components (e.g., $\sigma_v$~=~1,500 km s$^{-1}$ while the center radial speed $v_r$~=~4,000 km s$^{-1}$) and comparable to but less than the redshift for IGE components (e.g., typical $\sigma_v$~=~2,500 km s$^{-1}$ while the center line-of-sight speed $v_r$~=~2,500 -- 4,000 km s$^{-1}$). The fact that the fitted broadening $\sigma_v$ does not decrease with angular distance from the center, suggests that it is dominated by non-radial motions and/or thermal broadening.

% =========================================================================================
\section{Conclusions}

The microcalorimeter array of XRISM/Resolve has observed Tycho's SNR, providing  unprecedented energy resolution for spectroscopy across portions of the remnant.
We find that the spectra of the five publicly available observations of Tycho's SNR can collectively be well described by NEI models of separate intermediate-mass element and iron-group element plasmas in spherical expansion, with some difference between front and back shells for iron-group elements. This is done for single sets of elemental abundances across the remnant, which can be compared to bulk nucleosynthesis results in supernova calculations. While the SNR clearly has kinematic variations on small scales, on larger scales it displays quasi-spherical symmetry. We see this in iron as well as silicon, the latter of which has previously been described \citep{Williams17}. Additional Resolve observations should further clarify the explosion and its interactions with the external medium.
% We do expect the recently taken SE observation to show some degree of asphericity
% This analysis could be done with other observatories

\begin{acknowledgments}

This research has made use of data obtained through the High Energy Astrophysics Science Archive Research Center online service, provided by the NASA/Goddard Space Flight Center.
We thank Clemson University for use of the Palmetto 2 high-performance computing cluster which aided us greatly in reducing computation time for our modeling.

\end{acknowledgments}

\begin{contribution}

TEH did the data processing, performed the spectral fits, and did a large amount of the data analysis.
MDL formulated the initial research concept, supervised this research, and assisted with data interpretation.
Both authors contributed equally to the writing of the manuscript's text.

\end{contribution}

\vspace{5mm}
\facilities{XRISM (Resolve), Chandra (ACIS)}

\software{SAOImageDS9 \citep{ds9}, matplotlib \citep{matplotlib}}

\clearpage
\appendix

\section{XSPEC Outputs} \label{appx:xspec}

In Table \ref{tab:xspec_abund} below, we show the abundance outputs from our fitting with the 90\% errors from our MCMC runs. On the following page in Table \ref{tab:NW_fit}, we show the same for all of the other fit parameters in the NE Radial fit shown in Figure~\ref{fig:NE_individual_spectra}.

\begin{deluxetable*}{cccccc}[hbt]
    \tablewidth{0pt}
    \tablecaption{XSPEC abundances from the five joint spectral fits.} %with 90\% errors.
    \tablehead{
        \colhead{Abundance wrt Solar} & \colhead{All Data} & \colhead{NW Radial} & \colhead{NE Radial} & \colhead{North Radial} & \colhead{South Radial}
    }
    \startdata
        Si / C & $66.0_{-3.2}^{+4.2}$ & $38.7_{-7.3}^{+3.2}$ & $82.7_{-4.6}^{+5.8}$ & $54.9_{-7.9}^{+5.0}$ & $48.5_{-5.3}^{+4.2}$ \\
        S / C  & $54.1_{-2.5}^{+3.6}$ & $31.5_{-5.8}^{+2.7}$ & $62.5_{-3.6}^{+4.2}$ & $40.7_{-5.7}^{+4.1}$ & $35.2_{-4.0}^{+3.0}$ \\
        Ar / C & $49.7_{-2.8}^{+4.1}$ & $28.8_{-5.5}^{+2.7}$ & $50.0_{-4.4}^{+4.3}$ & $33.8_{-4.8}^{+4.6}$ & $27.0_{-3.6}^{+3.3}$ \\
        Ca / C & $86.5_{-5.1}^{+6.9}$ & $47.6_{-9.3}^{+4.1}$ & $78.6_{-8.1}^{+6.6}$ & $47.3_{-6.0}^{+8.2}$ & $38.5_{-5.9}^{+4.6}$ \\
        Cr / C & $55.9_{-18.3}^{+15.2}$ & $31.0_{-13.7}^{+9.9}$ & $33.8_{-9.6}^{+14.9}$ & $12.4_{-5.9}^{+17.5}$ & $8.2_{-2.1}^{+2.5}$ \\
        Mn / C & $67.4_{-18.0}^{+29.4}$ & $28.0_{-18.5}^{+15.6}$ & $38.9_{-18.6}^{+24.6}$ & $25.3_{-17.6}^{+12.6}$ & $8.5_{-6.4}^{+7.0}$ \\
        Fe$_{\text{\,IGE}}$ / C & $48.7_{-4.5}^{+2.8}$ & $23.8_{-5.9}^{+1.8}$ & $31.0_{-4.2}^{+2.3}$ & $18.1_{-2.2}^{+3.5}$ & $13.9_{-2.8}^{+1.4}$ \\
        Fe$_{\text{\,IME}}$ / C & $\le4.9$ & $\le1.8$ & $\le1.9$ & $\le1.1$ & $\le0.7$ \\
        Ni / C & $5.1_{-2.8}^{+2.6}$ & $6.1_{-4.4}^{+3.3}$ & $\le5.6$ & $2.4_{-1.2}^{+3.6}$ & $2.6_{-1.3}^{+2.3}$ \\
    \enddata
    \label{tab:xspec_abund}
    \tablecomments{All fits include the center spectrum and the column headings are which additional spectra it is fit simultaneously with.} % Raw XSPEC abundances should not be taken literally as they are dependent on the normalizations of the components
\end{deluxetable*}

\begin{deluxetable*}{cccccc}[htb] % this seems like a good way to display the data, open to suggestions though. If we lived in 4 spacial dimensions, this would look lovely as a 3D table.
    \tablewidth{0pt}
    \tablecaption{Non-abundance fit parameters for the NE radial fit. Column headings are the names of the spectra and IME/IGE component names are based on their colors in \Cref{fig:NE_individual_spectra}.}
    \tablehead{
        \colhead{Component} & \colhead{Parameter} & \colhead{Center Full} & \colhead{NE Dim} & \colhead{NE Bright} & \colhead{NE Shock}
    }
    \startdata
        & kT (keV) & $1.78_{-0.57}^{+0.15}$ & $5.43_{-2.61}^{+1.70}$ & $1.06_{-0.36}^{+0.19}$ & $4.22_{-0.94}^{+1.95}$ \\
        & $\tau$ ($10^{10}$ s/cm$^{3}$) & $3.59_{-0.30}^{+1.75}$ & $2.29_{-0.35}^{+0.67}$ & $7.72_{-1.82}^{+10.97}$ & $4.37_{-0.77}^{+1.37}$ \\
        IME Orange & z ($10^{-3}$) & $13.68_{-1.47}^{+0.76}$ & $8.08^{\,b}$ & $4.63^{\,b}$ & $2.20^{\,b}$ \\
        & $\sigma_v$ (km/s) & $1492_{-132}^{+568}$ & $1288_{-255}^{+315}$ & $2817_{-455}^{+628}$ & $2956_{-311}^{+467}$ \\
        & norm ($10^{-1}$) & $6.35_{-1.34}^{+1.41}$ & $2.56_{-0.87}^{+0.58}$ & $9.34_{-2.73}^{+5.41}$ & $4.24_{-1.56}^{+0.74}$ \\
        \hline
        & kT (keV) & $2.61_{-0.55}^{+0.40}$ & $1.66_{-0.25}^{+0.23}$ & $2.59_{-0.64}^{+0.56}$ & $1.68_{-0.19}^{+0.17}$ \\
        & $\tau$ ($10^{10}$ s/cm$^{3}$) & $2.40_{-0.19}^{+0.30}$ & $3.98_{-0.53}^{+0.99}$ & $2.82_{-0.34}^{+0.65}$ & $3.18_{-0.32}^{+0.54}$ \\
        IME Cyan & z ($10^{-3}$) & $-13.08_{-0.24}^{+0.36}$ & $-7.72^{\,b}$ & $-4.42^{\,b}$ & $-2.10^{\,b}$ \\
        & $\sigma_v$ (km/s) & $1217_{-67}^{+85}$ & $1614_{-111}^{+180}$ & $1916_{-187}^{+181}$ & $1432_{-120}^{+205}$ \\
        & norm ($10^{-1}$) & $7.81_{-1.09}^{+1.35}$ & $13.5_{-2.3}^{+2.7}$ & $10.0_{-0.5}^{+0.5}$ & $16.2_{-3.0}^{+4.0}$ \\
        \hline
        & kT (keV) & $10.0^{\,a}$ & $10.0^{\,a}$ & $10.0^{\,a}$ & $10.0^{\,a}$ \\
        & $\tau$ ($10^{10}$ s/cm$^{3}$) & $1.30_{-0.07}^{+0.62}$ & $0.76_{-0.18}^{+0.53}$ & $0.92_{-0.18}^{+0.46}$ & $1.12_{-0.15}^{+0.32}$ \\
        IGE Red & z ($10^{-3}$) & $12.34_{-2.25}^{+1.59}$ & $7.28^{\,b}$ & $4.17^{\,b}$ & $1.99^{\,b}$ \\
        & $\sigma_v$ (km/s) & $1935_{-541}^{+513}$ & $2713_{-807}^{+379}$ & $2495_{-380}^{+355}$ & $1922_{-454}^{+781}$ \\
        & norm ($10^{-1}$) & $2.07_{-0.35}^{+0.29}$ & $2.10_{-0.53}^{+0.45}$ & $2.40_{-0.45}^{+0.58}$ & $1.46_{-0.27}^{+0.78}$ \\
        \hline
        & kT (keV) & $6.14_{-0.90}^{+1.55}$ & $6.80_{-0.86}^{+2.67}$ & $10.0^{\,a}$ & $10.0^{\,a}$ \\
        & $\tau$ ($10^{10}$ s/cm$^{3}$) & $1.55_{-0.13}^{+0.13}$ & $1.26_{-0.21}^{+0.34}$ & $1.18_{-0.33}^{+0.30}$ & $0.77_{-0.38}^{+0.25}$ \\
        IGE Blue & z ($10^{-3}$) & $-7.75_{-0.32}^{+0.49}$ & $-4.57^{\,b}$ & $-2.62^{\,b}$ & $-1.25^{\,b}$ \\
        & $\sigma_v$ (km/s) & $946_{-90}^{+219}$ & $3895_{-832}^{+441}$ & $2255_{-950}^{+845}$ & $3485_{-1704}^{+3422}$ \\
        & norm ($10^{-1}$) & $2.28_{-0.46}^{+0.51}$ & $2.97_{-0.84}^{+0.92}$ & $1.19_{-0.48}^{+0.53}$ & $1.08_{-0.66}^{+0.52}$ \\
        \hline
        \multirow{2}{*}{Power-Law} & $\Gamma$ & $2.87_{-0.06}^{+0.06}$ & = Center & = Center & = Center \\
        & norm ($10^{-1}$) & $0.79_{-0.07}^{+0.09}$ & $0.74_{-0.07}^{+0.09}$ & $0.79_{-0.08}^{+0.08}$ & $1.01_{-0.11}^{+0.11}$ \\
        \hline
        NXB & amplitude & $0.83_{-0.07}^{+0.09}$ & $0.36_{-0.08}^{+0.06}$ & $0.32_{-0.06}^{+0.06}$ & $0.37_{-0.06}^{+0.08}$ \\
        \hline
        \multicolumn{2}{c}{C-Stat/Bins} & $3304/2997$ & $2557/2633$ & $2732/2845$ & $2687/2783$ \\
    \enddata
    \tablecomments{$^a$Stuck at parameter's prescribed hard upper limit \\ 
    $^b$Fixed to z$_\text{center}\,*\,$scaling values in rightmost column of Table~\ref{tab:geomeans}}
    \label{tab:NW_fit}
\end{deluxetable*}

\clearpage %%% temporary to push this all down a page

\bibliography{xrism_radial}{}

@ARTICLE{SatoHughes17,
       author = {{Sato}, Toshiki and {Hughes}, John P.},
        title = "{Direct Ejecta Velocity Measurements of Tycho's Supernova Remnant}",
      journal = {\apj},
         year = 2017,
        month = may,
       volume = {840},
       number = {2},
          eid = {112},
        pages = {112},
          doi = {10.3847/1538-4357/aa6f60},
archivePrefix = {arXiv},
       eprint = {1605.09059},
 primaryClass = {astro-ph.HE},
       adsurl = {https://ui.adsabs.harvard.edu/abs/2017ApJ...840..112S}
}

@ARTICLE{Williams17,
       author = {{Williams}, Brian J. and {Coyle}, Nina M. and {Yamaguchi}, Hiroya and {Depasquale}, Joseph and {Seitenzahl}, Ivo R. and {Hewitt}, John W. and {Blondin}, John M. and {Borkowski}, Kazimierz J. and {Ghavamian}, Parviz and {Petre}, Robert and {Reynolds}, Stephen P.},
        title = "{The Three-dimensional Expansion of the Ejecta from Tycho's Supernova Remnant}",
      journal = {\apj},
         year = 2017,
        month = jun,
       volume = {842},
       number = {1},
          eid = {28},
        pages = {28},
          doi = {10.3847/1538-4357/aa7384},
archivePrefix = {arXiv},
       eprint = {1705.05405},
 primaryClass = {astro-ph.HE},
       adsurl = {https://ui.adsabs.harvard.edu/abs/2017ApJ...842...28W}
}

@ARTICLE{Tsunemi86,
       author = {{Tsunemi}, H. and {Yamashita}, K. and {Masai}, K. and {Hayakawa}, S. and {Koyama}, K.},
        title = "{X-Ray Spectra of the Cassiopeia A and TYCHO Supernova Remnants and Their Element Abundances}",
      journal = {\apj},
         year = 1986,
        month = jul,
       volume = {306},
        pages = {248},
          doi = {10.1086/164337},
       adsurl = {https://ui.adsabs.harvard.edu/abs/1986ApJ...306..248T}
}

@ARTICLE{Katsuda2010,
       author = {{Katsuda}, Satoru and {Petre}, Robert and {Hughes}, John P. and {Hwang}, Una and {Yamaguchi}, Hiroya and {Hayato}, Asami and {Mori}, Koji and {Tsunemi}, Hiroshi},
        title = "{X-ray Measured Dynamics of Tycho's Supernova Remnant}",
      journal = {\apj},
         year = 2010,
        month = feb,
       volume = {709},
       number = {2},
        pages = {1387-1395},
          doi = {10.1088/0004-637X/709/2/1387},
archivePrefix = {arXiv},
       eprint = {1001.2484},
 primaryClass = {astro-ph.HE},
       adsurl = {https://ui.adsabs.harvard.edu/abs/2010ApJ...709.1387K}
}

@ARTICLE{GoodmanWeare,
       author = {{Goodman}, Jonathan and {Weare}, Jonathan},
        title = "{Ensemble samplers with affine invariance}",
      journal = {Communications in Applied Mathematics and Computational Science},
         year = 2010,
        month = jan,
       volume = {5},
       number = {1},
        pages = {65-80},
          doi = {10.2140/camcos.2010.5.65},
       adsurl = {https://ui.adsabs.harvard.edu/abs/2010CAMCS...5...65G}
}

@ARTICLE{Leahy2024,
       author = {{Leahy}, D.~A. and {Foster}, A. and {Seitenzahl}, I.},
        title = "{On the Interpretation of XSPEC Abundances and Emission Measures}",
      journal = {\aj},
         year = 2024,
        month = mar,
       volume = {167},
       number = {3},
          eid = {130},
        pages = {130},
          doi = {10.3847/1538-3881/ad2359},
       adsurl = {https://ui.adsabs.harvard.edu/abs/2024AJ....167..130L}
}

@ARTICLE{Borkowski2001,
       author = {{Borkowski}, Kazimierz J. and {Lyerly}, William J. and {Reynolds}, Stephen P.},
        title = "{Supernova Remnants in the Sedov Expansion Phase: Thermal X-Ray Emission}",
      journal = {\apj},
         year = 2001,
        month = feb,
       volume = {548},
       number = {2},
        pages = {820-835},
          doi = {10.1086/319011},
archivePrefix = {arXiv},
       eprint = {astro-ph/0008066},
 primaryClass = {astro-ph},
       adsurl = {https://ui.adsabs.harvard.edu/abs/2001ApJ...548..820B}
}

@ARTICLE{Wilms2000,
       author = {{Wilms}, J. and {Allen}, A. and {McCray}, R.},
        title = "{On the Absorption of X-Rays in the Interstellar Medium}",
      journal = {\apj},
         year = 2000,
        month = oct,
       volume = {542},
       number = {2},
        pages = {914-924},
          doi = {10.1086/317016},
archivePrefix = {arXiv},
       eprint = {astro-ph/0008425},
 primaryClass = {astro-ph},
       adsurl = {https://ui.adsabs.harvard.edu/abs/2000ApJ...542..914W}
}

@ARTICLE{Fujimaru26,
       author = {{Fujimaru}, Yusei and {Lee}, Shiu-Hang and {Ferrand}, Gilles and {Patnaude}, Daniel and {Nagataki}, Shigehiro and {Pakmor}, R{\"u}diger and {Safi-Harb}, Samar and {R{\"o}pke}, Friedrich K. and {Decourchelle}, Anne and {Seitenzahl}, Ivo R.},
        title = "{Probing the Diversity of Type Ia Supernova Remnants in 3D Hydrodynamic Simulations with X-Ray Spectral Synthesis}",
      journal = {\apj},
         year = 2026,
        month = mar,
       volume = {999},
       number = {2},
          eid = {151},
        pages = {151},
          doi = {10.3847/1538-4357/ae4004},
archivePrefix = {arXiv},
       eprint = {2601.20446},
 primaryClass = {astro-ph.HE},
       adsurl = {https://ui.adsabs.harvard.edu/abs/2026ApJ...999..151F}
}

@ARTICLE{Godinaud2025,
       author = {{Godinaud}, L. and {Acero}, F. and {Decourchelle}, A. and {Ballet}, J.},
        title = "{Bayesian insights into the Tycho supernova remnant: A detailed mapping of ejecta properties}",
      journal = {\aap},
         year = 2025,
        month = jan,
       volume = {693},
          eid = {A234},
        pages = {A234},
          doi = {10.1051/0004-6361/202450518},
archivePrefix = {arXiv},
       eprint = {2404.17296},
 primaryClass = {astro-ph.HE},
       adsurl = {https://ui.adsabs.harvard.edu/abs/2025A&A...693A.234G}
}

@ARTICLE{Vancura95,
       author = {{Vancura}, O. and {Gorenstein}, P. and {Hughes}, John P.},
        title = "{Evidence for Elemental Variation in the Ejecta of the TYCHO Supernova Remnant}",
      journal = {\apj},
         year = 1995,
        month = mar,
       volume = {441},
        pages = {680},
          doi = {10.1086/175391},
       adsurl = {https://ui.adsabs.harvard.edu/abs/1995ApJ...441..680V}
}

@ARTICLE{KaastraBleeker,
       author = {{Kaastra}, J.~S. and {Bleeker}, J.~A.~M.},
        title = "{Optimal binning of X-ray spectra and response matrix design}",
      journal = {\aap},
         year = 2016,
        month = mar,
       volume = {587},
          eid = {A151},
        pages = {A151},
          doi = {10.1051/0004-6361/201527395},
archivePrefix = {arXiv},
       eprint = {1601.05309},
 primaryClass = {astro-ph.IM},
       adsurl = {https://ui.adsabs.harvard.edu/abs/2016A&A...587A.151K}
}

@ARTICLE{Plucinsky2025,
       author = {{Plucinsky}, Paul and {Agarwal}, Manan and {Gu}, Liyi and {Foster}, Adam and {Sato}, Toshiki and {Bamba}, Aya and {Vink}, Jacco and {Ichihashi}, Masahiro and {Matsunaga}, Kai and {Mori}, Koji and {Nakajima}, Hiroshi and {Porter}, Frederick S and {Sonoda}, Haruto and {Suzuki}, Shunsuke and {Tateishi}, Dai and {Terada}, Yukikatsu and {Uchida}, Hiroyuki and {Yamaguchi}, Hiroya},
        title = "{XRISM observations of Cassiopeia A: Overview, atomic data, and spectral models}",
      journal = {\pasj},
         year = 2025,
        month = sep,
       volume = {77},
        pages = {S171-S187},
          doi = {10.1093/pasj/psaf091},
archivePrefix = {arXiv},
       eprint = {2508.01089},
 primaryClass = {astro-ph.HE},
       adsurl = {https://ui.adsabs.harvard.edu/abs/2025PASJ...77S.171P}
}

@INPROCEEDINGS{XSPEC,
       author = {{Arnaud}, K.~A.},
        title = "{XSPEC: The First Ten Years}",
    booktitle = {Astronomical Data Analysis Software and Systems V},
         year = 1996,
       editor = {{Jacoby}, George H. and {Barnes}, Jeannette},
       series = {Astronomical Society of the Pacific Conference Series},
       volume = {101},
        month = jan,
        pages = {17},
       adsurl = {https://ui.adsabs.harvard.edu/abs/1996ASPC..101...17A}
}

@ARTICLE{Pravdo80,
       author = {{Pravdo}, S.~H. and {Smith}, B.~W. and {Charles}, P.~A. and {Tuohy}, I.~R.},
        title = "{X-ray line emission from the TYCHO supernova remnant}",
      journal = {\apjl},
         year = 1980,
        month = jan,
       volume = {235},
        pages = {L9-L12},
          doi = {10.1086/183147},
       adsurl = {https://ui.adsabs.harvard.edu/abs/1980ApJ...235L...9P}
}

@ARTICLE{Seward83,
       author = {{Seward}, F. and {Gorenstein}, P. and {Tucker}, W.},
        title = "{The mass of Tycho's supernova remnant as determined from a high-resolution X-ray map.}",
      journal = {\apj},
         year = 1983,
        month = mar,
       volume = {266},
        pages = {287-297},
          doi = {10.1086/160778},
       adsurl = {https://ui.adsabs.harvard.edu/abs/1983ApJ...266..287S}
}

@ARTICLE{Hwang98_asca,
       author = {{Hwang}, Una and {Hughes}, John P. and {Petre}, Robert},
        title = "{The X-Ray Iron Emission from Tycho's Supernova Remnant}",
      journal = {\apj},
         year = 1998,
        month = apr,
       volume = {497},
       number = {2},
        pages = {833-841},
          doi = {10.1086/305501},
archivePrefix = {arXiv},
       eprint = {astro-ph/9712241},
 primaryClass = {astro-ph},
       adsurl = {https://ui.adsabs.harvard.edu/abs/1998ApJ...497..833H}
}

@ARTICLE{Martinez-Rodriguez18,
       author = {{Mart{\'\i}nez-Rodr{\'\i}guez}, H{\'e}ctor and {Badenes}, Carles and {Lee}, Shiu-Hang and {Patnaude}, Daniel J. and {Foster}, Adam R. and {Yamaguchi}, Hiroya and {Auchettl}, Katie and {Bravo}, Eduardo and {Slane}, Patrick O. and {Piro}, Anthony L. and {Park}, Sangwook and {Nagataki}, Shigehiro},
        title = "{Chandrasekhar and Sub-Chandrasekhar Models for the X-Ray Emission of Type Ia Supernova Remnants. I. Bulk Properties}",
      journal = {\apj},
         year = 2018,
        month = oct,
       volume = {865},
       number = {2},
          eid = {151},
        pages = {151},
          doi = {10.3847/1538-4357/aadaec},
archivePrefix = {arXiv},
       eprint = {1807.02990},
 primaryClass = {astro-ph.HE},
       adsurl = {https://ui.adsabs.harvard.edu/abs/2018ApJ...865..151M}
}

@ARTICLE{Davison76,
       author = {{Davison}, P.~J.~N. and {Culhane}, J.~L. and {Mitchell}, R.~J.},
        title = "{X-ray spectra of Cassiopeia A and Tycho's supernova observed with Ariel-5.}",
      journal = {\apjl},
         year = 1976,
        month = may,
       volume = {206},
        pages = {L37-L40},
          doi = {10.1086/182127},
       adsurl = {https://ui.adsabs.harvard.edu/abs/1976ApJ...206L..37D}
}

@ARTICLE{Gorenstein70,
       author = {{Gorenstein}, P. and {Kellogg}, E.~M. and {Gursky}, H.},
        title = "{X-Ray Characteristics of Three Supernova Remnants}",
      journal = {\apj},
         year = 1970,
        month = apr,
       volume = {160},
        pages = {199},
          doi = {10.1086/150417},
       adsurl = {https://ui.adsabs.harvard.edu/abs/1970ApJ...160..199G}
}

@ARTICLE{Decourchelle01,
       author = {{Decourchelle}, A. and {Sauvageot}, J.~L. and {Audard}, M. and {Aschenbach}, B. and {Sembay}, S. and {Rothenflug}, R. and {Ballet}, J. and {Stadlbauer}, T. and {West}, R.~G.},
        title = "{XMM-Newton observation of the Tycho supernova remnant}",
      journal = {\aap},
         year = 2001,
        month = jan,
       volume = {365},
        pages = {L218-L224},
          doi = {10.1051/0004-6361:20000115},
archivePrefix = {arXiv},
       eprint = {astro-ph/0012288},
 primaryClass = {astro-ph},
       adsurl = {https://ui.adsabs.harvard.edu/abs/2001A&A...365L.218D}
}

@ARTICLE{Baade45,
       author = {{Baade}, W.},
        title = "{B Cassiopeiae as a Supernova of Type I.}",
      journal = {\apj},
         year = 1945,
        month = nov,
       volume = {102},
        pages = {309},
          doi = {10.1086/144761},
       adsurl = {https://ui.adsabs.harvard.edu/abs/1945ApJ...102..309B}
}

@ARTICLE{Krause08,
       author = {{Krause}, Oliver and {Tanaka}, Masaomi and {Usuda}, Tomonori and {Hattori}, Takashi and {Goto}, Miwa and {Birkmann}, Stephan and {Nomoto}, Ken'ichi},
        title = "{Tycho Brahe's 1572 supernova as a standard type Ia as revealed by its light-echo spectrum}",
      journal = {\nat},
         year = 2008,
        month = dec,
       volume = {456},
       number = {7222},
        pages = {617-619},
          doi = {10.1038/nature07608},
archivePrefix = {arXiv},
       eprint = {0810.5106},
 primaryClass = {astro-ph},
       adsurl = {https://ui.adsabs.harvard.edu/abs/2008Natur.456..617K}
}

@ARTICLE{Becker80,
       author = {{Becker}, R.~H. and {Smith}, B.~W. and {White}, N.~E. and {Mushotzky}, R.~F. and {Holt}, S.~S. and {Boldt}, E.~A. and {Serlemitsos}, P.~J.},
        title = "{Elemental abundances in a Type I supernova remnant}",
      journal = {\apjl},
         year = 1980,
        month = jan,
       volume = {235},
        pages = {L5-L8},
          doi = {10.1086/183146},
       adsurl = {https://ui.adsabs.harvard.edu/abs/1980ApJ...235L...5B}
}

@ARTICLE{Arnett69,
       author = {{Arnett}, W. David},
        title = "{A Possible Model of Supernovae: Detonation of $^{12}$C}",
      journal = {\apss},
         year = 1969,
        month = oct,
       volume = {5},
       number = {2},
        pages = {180-212},
          doi = {10.1007/BF00650291},
       adsurl = {https://ui.adsabs.harvard.edu/abs/1969Ap&SS...5..180A}
}

@ARTICLE{Nomoto1984,
       author = {{Nomoto}, K. and {Thielemann}, F.-K. and {Yokoi}, K.},
        title = "{Accreting white dwarf models for type I supernovae. III. Carbon deflagration supernovae.}",
      journal = {\apj},
         year = 1984,
        month = nov,
       volume = {286},
        pages = {644-658},
          doi = {10.1086/162639},
       adsurl = {https://ui.adsabs.harvard.edu/abs/1984ApJ...286..644N}
}

@ARTICLE{Nomoto1997,
       author = {{Nomoto}, K. and {Iwamoto}, K. and {Nakasato}, N. and {Thielemann}, F.-K. and {Brachwitz}, F. and {Tsujimoto}, T. and {Kubo}, Y. and {Kishimoto}, N.},
        title = "{Nucleosynthesis in type Ia supernovae}",
      journal = {\nphysa},
         year = 1997,
        month = feb,
       volume = {621},
        pages = {467-476},
          doi = {10.1016/S0375-9474(97)00291-1},
archivePrefix = {arXiv},
       eprint = {astro-ph/9706025},
 primaryClass = {astro-ph},
       adsurl = {https://ui.adsabs.harvard.edu/abs/1997NuPhA.621..467N}
}

@ARTICLE{PyAtomDB,
       author = {{Foster}, Adam R. and {Heuer}, Keri},
        title = "{PyAtomDB: Extending the AtomDB Atomic Database to Model New Plasma Processes and Uncertainties}",
      journal = {Atoms},
         year = 2020,
        month = aug,
       volume = {8},
       number = {3},
        pages = {49},
          doi = {10.3390/atoms8030049},
       adsurl = {https://ui.adsabs.harvard.edu/abs/2020Atoms...8...49F}
}

@ARTICLE{Hamilton86,
       author = {{Hamilton}, A.~J.~S. and {Sarazin}, C.~L. and {Szymkowiak}, A.~E.},
        title = "{The X-Ray Spectrum of TYCHO}",
      journal = {\apj},
         year = 1986,
        month = jan,
       volume = {300},
        pages = {713},
          doi = {10.1086/163845},
       adsurl = {https://ui.adsabs.harvard.edu/abs/1986ApJ...300..713H}
}

@ARTICLE{XRISM,
       author = {{Tashiro}, Makoto and {Kelley}, Richard and {Watanabe}, Shin and {Maejima}, Hironori and {Reichenthal}, Lillian and {Toda}, Kenichi and {Hartz}, Leslie and {Santovincenzo}, Andrea and {Matsushita}, Kyoko and {Yamaguchi}, Hiroya and {Petre}, Robert and {Williams}, Brian and {Guainazzi}, Matteo and {Costantini}, Elisa and {Takei}, Yoh and {Ishisaki}, Yoshitaka and {Fujimoto}, Ryuichi and {Henegar-Leon}, Joy and {Sneiderman}, Gary and {Tomida}, Hiroshi and {Mori}, Koji and {Nakajima}, Hiroshi and {Terada}, Yukikatsu and {Holland}, Matthew and {Loewenstein}, Michael and {Miller}, Eric and {Sawada}, Makoto and {Kallman}, Timothy and {Kaastra}, Jelle and {Done}, Chris and {Enoto}, Teruaki and {Bamba}, Aya and {Corrales}, Lia and {Ueda}, Yoshihiro and {Kara}, Erin and {Zhuravleva}, Irina and {Fujita}, Yutaka and {Arai}, Yoshitaka and {Audard}, Marc and {Awaki}, Hisamitsu and {Ballhausen}, Ralf and {Baluta}, Chris and {Bando}, Nobutaka and {Behar}, Ehud and {Bialas}, Thomas and {Boissay-Malaquin}, Rozenn and {Brenneman}, Laura and {Brown}, Gregory V. and {Chiao}, Meng and {Cumbee}, Renata and {de Vries}, Cor and {den Herder}, Jan-Willem and {D{\'\i}az Trigo}, Mar{\'\i}a and {DiPirro}, Michael and {Dotani}, Tadayasu and {Carrero}, Jacobo Ebrero and {Ebisawa}, Ken and {Eckart}, Megan and {Eckert}, Dominique and {Eguchi}, Satoshi and {Ezoe}, Yuichiro and {Ferrigno}, Carlo and {Foster}, Adam and {Fukazawa}, Yasushi and {Fukushima}, Kotaro and {Furuzawa}, Akihiro and {Gallo}, Luigi C. and {Garcia Martinez}, Javier and {Gorter}, Nathalie and {Grim}, Martin and {Gu}, Liyi and {Hagino}, Kouichi and {Hamaguchi}, Kenji and {Hatsukade}, Isamu and {Hayashi}, Katsuhiro and {Hayashi}, Takayuki and {Hell}, Natalie and {Hodges-Kluck}, Edmund and {Horiuchi}, Takafumi and {Hornschemeier}, Ann and {Hoshino}, Akio and {Ichinohe}, Yuto and {Ikuta}, Chisato and {Iizuka}, Ryo and {Ishi}, Daiki and {Ishida}, Manabu and {Ishihama}, Naoki and {Ishikawa}, Kumi and {Ishimura}, Kosei and {Jaffe}, Tess and {Katsuda}, Satoru and {Kanemaru}, Yoshiaki and {Kenyon}, Steven and {Kilbourne}, Caroline and {Kimball}, Mark and {Kitamoto}, Shunji and {Kobayashi}, Shogo and {Kohmura}, Takayoshi and {Kubota}, Aya and {Leutenegger}, Maurice A. and {Maeda}, Yoshitomo and {Markevitch}, Maxim and {Matsumoto}, Hironori and {Matsuzaki}, Keiichi and {McCammon}, Dan and {McLaughlin}, Brian and {McNamara}, Brian and {Mernier}, Fran{\c{c}}ois and {Miko}, Joseph and {Miller}, Jon M. and {Minesugi}, Kenji and {Mitani}, Shinji and {Mitsuishi}, Ikuyuki and {Mizumoto}, Misaki and {Mizuno}, Tsunefumi and {Mukai}, Koji and {Murakami}, Hiroshi and {Mushotzky}, Richard and {Nakazawa}, Kazuhiro and {Natsukari}, Chikara and {Ness}, Jan-Uwe and {Nigo}, Kenichiro and {Nishiyama}, Mari and {Nobukawa}, Kumiko and {Nobukawa}, Masayoshi and {Noda}, Hirofumi and {Odaka}, Hirokazu and {Ogawa}, Mina and {Ogawa}, Shoji and {Ogorzalek}, Anna and {Okajima}, Takashi and {Okamoto}, Atsushi and {Ota}, Naomi and {Ozaki}, Masanobu and {Paltani}, Stephane and {Plucinsky}, Paul and {Porter}, F. Scott and {Pottschmidt}, Katja and {Quero}, Jose Antonio and {Sasaki}, Takahiro and {Sato}, Kosuke and {Sato}, Rie and {Sato}, Toshiki and {Sato}, Yoichi and {Seta}, Hiromi and {Shida}, Maki and {Shidatsu}, Megumi and {Shigeto}, Shuhei and {Shipman}, Russel and {Shinozaki}, Keisuke and {Shirron}, Peter and {Simionescu}, Aurora and {Smith}, Randall K. and {Soong}, Yang and {Suzuki}, Hiromasa and {Szymkowiak}, Andrew and {Takahashi}, Hiromitsu and {Takeo}, Mai and {Tamagawa}, Toru and {Tamura}, Keisuke and {Tanaka}, Takaaki and {Tanimoto}, Atsushi and {Terashima}, Yuichi and {Tsuboi}, Yohko and {Tsujimoto}, Masahiro and {Tsunemi}, Hiroshi and {Tsuru}, Takeshi Go and {Uchida}, Hiroyuki and {Uchida}, Nagomi and {Uchida}, Yuusuke and {Uchiyama}, Hideki and {Uno}, Shinichiro and {Vink}, Jacco and {Witthoeft}, Michael and {Wolfs}, Rob and {Yamada}, Satoshi and {Yamada}, Shinya and {Yamaoka}, Kazutaka and {Yamasaki}, Noriko and {Yamauchi}, Makoto and {Yamauchi}, Shigeo and {Yanagase}, Keiichi and {Yaqoob}, Tahir and {Yasuda}, Susumu and {Yoneyama}, Tomokage and {Yoshida}, Tessei and {Yukita}, Mihoko},
        title = "{X-Ray Imaging and Spectroscopy Mission}",
      journal = {\pasj},
         year = 2025,
        month = sep,
       volume = {77},
        pages = {S1-S9},
          doi = {10.1093/pasj/psaf023},
       adsurl = {https://ui.adsabs.harvard.edu/abs/2025PASJ...77S...1T}
}

@INPROCEEDINGS{ds9,
       author = {{Joye}, W.~A. and {Mandel}, E.},
        title = "{New Features of SAOImage DS9}",
    booktitle = {Astronomical Data Analysis Software and Systems XII},
         year = 2003,
       editor = {{Payne}, H.~E. and {Jedrzejewski}, R.~I. and {Hook}, R.~N.},
       series = {Astronomical Society of the Pacific Conference Series},
       volume = {295},
        month = jan,
        pages = {489},
       adsurl = {https://ui.adsabs.harvard.edu/abs/2003ASPC..295..489J}
}

@ARTICLE{matplotlib,
       author = {{Hunter}, John D.},
        title = "{Matplotlib: A 2D Graphics Environment}",
      journal = {Computing in Science and Engineering},
         year = 2007,
        month = jan,
       volume = {9},
       number = {3},
        pages = {90-95},
          doi = {10.1109/MCSE.2007.55},
       adsurl = {https://ui.adsabs.harvard.edu/abs/2007CSE.....9...90H}
}
\bibliographystyle{aasjournalv7.1}

\end{document}